\documentclass[a4paper,11pt]{article}
\pdfoutput=1 %

\usepackage{jheppub} 
\usepackage{amsmath,amssymb,graphicx,float,slashed,xcolor,multicol}
\usepackage{tabularx}
\usepackage{url}
\usepackage{footmisc}
\usepackage{amsfonts}
\usepackage{cancel}
\usepackage{color}
\usepackage{multirow} 
\usepackage{pifont}
\usepackage{epstopdf}
\usepackage{comment}
\usepackage{booktabs} 
\usepackage{natbib}
\usepackage{array}
\usepackage{mathrsfs}
\usepackage[toc,page]{appendix}
\usepackage{mathtools}
\usepackage{romannum}
\usepackage{ulem}
\usepackage{bbold}
\usepackage{enumitem}
\usepackage{multirow}
\usepackage{cleveref}
\usepackage{caption,subcaption}
\usepackage{float}
\usepackage{multirow}
\usepackage{upgreek}
\usepackage[toc,page]{appendix}
\usepackage{romannum}
\allowdisplaybreaks
\usepackage{bbm}                
\usepackage{xspace}				
\usepackage{pgffor}							

\newcommand*{\rom}[1]{\expandafter\@slowromancap\romannumeral #1@}

\newcommand{\RNum}[1]{\uppercase\expandafter{\romannumeral #1\relax}}

\def\lag{\mathscr{L}}

\def\ra{\rightarrow}

\def\beq{\begin{equation}}
\def\eeq{\end{equation}}
\def\beqa{\begin{eqnarray}}
\def\eeqa{\end{eqnarray}} 
\title{Right-handed Dirac and Majorana neutrinos at Belle II}
\author[a]{Tao Han,}
\author[b]{Jiajun Liao,}
\author[c]{Hongkai Liu,}
\author[d,e]{Danny Marfatia}
\affiliation[a]{Department of Physics and Astronomy, University of Pittsburgh, Pittsburgh, PA 15260, USA }
\affiliation[b]{School of Physics, Sun Yat-Sen University, Guangzhou, 510275, China}
\affiliation[c]{Department of Physics, Technion – Israel Institute of Technology, Haifa 3200003, Israel}
\affiliation[d]{Department of Physics and Astronomy, University of Hawaii at Manoa, Honolulu, HI 96822, USA}
\affiliation[e]{Kavli Institute for Theoretical Physics, University of California, Santa Barbara, CA 93106, USA}
\emailAdd{than@pitt.edu}
\emailAdd{liaojiajun@mail.sysu.edu.cn}
\emailAdd{liu.hongkai@campus.technion.ac.il}
\emailAdd{dmarf8@hawaii.edu}

\preprint{
\begin{flushright}
PITT-PACC-2208
\end{flushright}
}

\abstract{
We assess the ability of the Belle~II experiment to probe the Dirac or Majorana nature of a massive right-handed neutrino (RHN) $N$ in the MeV to GeV mass range. We consider the production and decay of RHNs to proceed via new interactions described by the standard model effective field theory (SMEFT) extended with right-handed neutrino fields (SMNEFT), and not via mass mixing with active neutrinos. 
We find that Belle II has the potential to discover $N$ if kinematically accessible.
We perform detailed simulations of the angular distributions of lepton pairs from the decay of $N$ produced in two-body and three-body decays of $B$ mesons.
We show that for $m_N$ above 100~MeV, Belle~II can distinguish between Dirac and Majorana neutrinos
at more than the 5$\sigma$ CL for most operators, and the combination of the production and decay operators can be identified from the subsequent decay of the heavy neutrino. Also, the production operators can be identified using three-body $B$ meson decay for any $m_N$ if the $B\to D\ell N$ and $B\to D^*\ell N$ events can be well separated.
}

\begin{document}

\titlepage

\maketitle



\flushbottom

\section{Introduction}
\label{sec:intro}
Neutrino oscillation experiments and cosmological observations confirm that neutrino masses are nonzero, but tiny~\cite{Zyla:2020zbs}. Massive fermions can be either Dirac or Majorana particles. However, whether neutrinos are Dirac fermions (DFs) or Majorana fermions (MFs) is still an open question. If neutrinos are MFs, their masses can be naturally explained by the Type I seesaw mechanism, in which heavy Majorana neutrinos suppress the active neutrino masses. The determination of the nature of heavy neutrinos can also shed light on the neutrino mass mechanism. If confirmed, the existence of a MF would revolutionize our understanding of elementary particle physics.

The most promising approach to determine the nature of neutrinos is to search for lepton number violation (LNV), which has been under intensive theoretical and experimental investigation.
In this work, we make use of the RHN decay, $\bar N \ra \ell_\alpha^+ \ell_\beta^- \bar\nu$, to determine the nature of $N$. If $N$ is a MF, one could observe both $\ell^+_\alpha \ell^-_\beta$ and $\ell^+_\beta \ell^-_\alpha$ in the final state. However, if $\alpha = \beta$, the  LNV information is confined to the neutrino sector, and is undetectable at 
Belle~II. However, distinctive angular distributions from MF decay can serve as an alternative probe of the nature of neutrinos~\cite{Han:2012vk}.
Reference~\cite{Balantekin:2018ukw} proposed to use the kinematic distributions of two-body decays of the heavy neutrino $N$. If $N$ is a MF, there is no forward-backward asymmetry of the charged-particle-pair system in the $N$ rest frame. On the other hand, the distributions for a DF are less constrained and depend on the physics responsible for $N$ decay. Recently, it is has been pointed out that the kinematic distributions of three-body $N$ decays can also be used to determine the nature of heavy neutrinos, and probe the interaction structure of the heavy neutrino decay~\cite{deGouvea:2021ual}. 

A model-independent effective field theory (EFT) framework provides a natural description of the physics responsible for the production and decay of heavy neutrinos. The standard model effective field theory (SMEFT)~\cite{Grzadkowski:2010es,Henning:2014wua,Brivio:2017vri} is often used to explore new physics near or above the electroweak scale since it is composed of the standard model (SM) field content and respects the full SM gauge symmetry. In this work, we consider operators in the  SMNEFT framework~\cite{delAguila:2008ir, Aparici:2009fh, Liao:2016qyd, Bischer:2019ttk, Li:2021tsq}, which extends the SMEFT with right-handed neutrinos. 

It is known that a heavy Majorana neutrino can be searched for in meson decays, if kinematically accessible \cite{Atre:2009rg}. In this paper,
we focus on the discrimination of the Dirac versus Majorana nature of heavy neutrinos at the Belle II experiment, which provides high precision measurements of the decay products of the $B$ mesons~\cite{Belle-II:2018jsg}.  We consider the production of on-shell heavy
 antineutrinos $\bar N$ from both the leptonic two-body decay of $B$ mesons $B^- \ra \ell^- \bar N$ and the semi-leptonic three-body decay $\bar{B}^0 \ra D^{(*)+} \ell^- \bar N$.  The on-shell $\bar N$ decays to a same flavor charged lepton pair and a neutrino: $\bar N \ra \ell^+\ell^-\bar \nu$. 
The LHCb experiment has placed a strong bound on the branching fraction of $B^+ \to \mu^+\mu^-\mu^+ \nu_\mu$~\cite{LHCb:2018jvy}, which applies to $\ell =\mu$ in the two-body decay channel.
To evade this constraint and to open a wider window for $m_N$, we assume $\ell = e$ for both $B$ and $\bar N$ decays. Throughout, we consider a sterile neutrino $N$ that only couples to third generation quarks and first generation leptons. Complementary studies in the SMNEFT framework have considered $N$ produced at Belle~II to couple to light quarks or heavy leptons~\cite{Duarte:2019rzs, Duarte:2020vgj, Zhou:2021ylt,Beltran:2022ast}. 

The electron from $B$ decay can be singled out by using the $\bar N$ resonance peak, as $p_N^2 = m_N^2$ with $p_N = p_B  -p_e$ ($p_N = p_B - p_D -p_e$) in the two-body (three-body) decay of the $B$ meson. Belle~II will perform high-precision measurements of the energy and angular distributions of daughter charged particles from $\bar N$ decay, and will reach the sensitivity to probe the nature and interaction structure of heavy neutrinos by using the angular distributions of charged leptons. In this study, we use three benchmark masses, 1~GeV, 100~MeV, and 5~MeV. Kinematics limits Belle-II's ability to produce RHNs heavier than a few GeV, and for RHNs lighter than 5~MeV, the di-leptons from the decay of the highly boosted RHNs have a small opening angle so the angular distributions get smeared out due to the detector resolution. In the 5~MeV case, we will need to assume invisible decay modes of $\bar N$ to enforce a short lifetime and prompt decay. In some cases, $\bar N$ may have a decay length of $\sim 1$~cm, which leads to an identifiable displaced vertex. However, if there are enough hits in the central drift chamber, the decay can be treated as prompt despite the displaced vertex.

The paper is organized as follows. In Section~\ref{sec:FormalismB}, we discuss two- and three-body  decays of $B$ mesons in the SMNEFT framework. In Section~\ref{sec:FormalismN}, we discuss three-body decays of heavy antineutrinos $\bar N$. We present current bounds on the Wilson coefficients (WCs) in Section~\ref{sec:Bounds}. In Section~\ref{sec:Belle}, we study the angular distributions from $B$ and $\bar N$ decay at Belle~II, and the physics that can be extracted from them. We summarize our results in Section~\ref{sec:sum}.

\section{Production of $\bar N$ from $B$ meson decay}
\label{sec:FormalismB}

\begin{table}
	\centering
	\begin{tabular}{| c |  c |  c |  c | }
		\toprule
		 $\mathcal{O}^S_{RR}$ &  $\mathcal{O}^S_{LR}$&  $\mathcal{O}^V_{RR}$& $\mathcal{O}^T_{RR}$   \\
		\midrule
		  $\mathcal{O}^{(1)}_{LNQd}$ & $\mathcal{O}_{LNuQ}$&  $\mathcal{O}_{Nedu}$ & $\mathcal{O}^{(3)}_{LNQd}$   \\
		\toprule
	\end{tabular}
	\caption{The origin of low-energy effective operators from SMNEFT in the operator basis of Ref.~\cite{Liao:2016qyd}.}
	\label{Table: op}
\end{table}

$B$ meson decays to $\bar N$ can be described by the low-energy effective Lagrangian comprised of four-fermion contact interactions,
\beq
-\lag_{\text{eff}} = \frac{4G_FV_{cb}}{\sqrt{2}}\sum_{\substack{X=S,V,T\\ \alpha=L,R}}C^X_{\alpha R}~\mathcal{O}^X_{\alpha R}\,,
\label{eq:lag_B}
\eeq
where
\beqa
\mathcal{O}^V_{\alpha R} &\equiv& (\bar{u}\gamma^\mu P_\alpha b) (\bar{e}\gamma^\mu P_R N)\,,\\
\mathcal{O}^S_{\alpha R} &\equiv& (\bar{u} P_\alpha b) (\bar{e} P_R N)\,,\\
\mathcal{O}^T_{\alpha R} &\equiv& \delta_{\alpha R} (\bar{u} \sigma^{\mu\nu} P_\alpha b) (\bar{e}\sigma_{\mu\nu} P_R N)\,.
\eeqa
While these operators can not arise from SMEFT, $\mathcal{O}^V_{RR},\,\mathcal{O}^S_{LR},\,\mathcal{O}^S_{RR}$, and $\mathcal{O}^T_{RR}$ arise from SMNEFT; see Table~\ref{Table: op}. In this section, we focus on the three SMNEFT operators,
\beqa
\mathcal{O}_{LNuQ} &=& (\bar{L} N)(\bar{u}Q)\,,\nonumber\\
\label{eq:prod_opt}
\mathcal{O}_{Nedu} &=& (\bar{N} \gamma^{\mu} e)(\bar{d}\gamma_{\mu} u)\,,\\
\mathcal{O}^{(3)}_{LNQd} &=& (\overline{L^j}\sigma_{\mu\nu} N)\epsilon_{jk}(\overline{Q^k}\sigma^{\mu\nu}d)\,,\nonumber
\eeqa
where $L$ and $Q$  are left-handed SU(2) doublets, $e, u$ and $d$ are right-handed SU(2) singlets, and $N$ is the massive RHN. As stated earlier, we assume $N$ only couples to third generation quarks and first generation leptons.\footnote{Note that the scalar and tensor operators associated with heavy flavor can generate a large Dirac mass term $\bar N \nu_L$ at the two-loop level~\cite{Prezeau:2004md, Ito:2004sh} and contribute to the active neutrino masses; see Ref.~\cite{Han:2020pff} for order of magnitude estimates in SMNEFT. However, the bounds from neutrino masses are model dependent because of the possibility of fine-tuned cancellations arising from ultraviolet physics.} 

 No mixing between the three SMNEFT operators is introduced by gauge couplings~\cite{Datta:2020ocb}, and mixing introduced by Yukawa couplings is suppressed by the electron mass~\cite{Datta:2021akg}. 
Below the weak scale, a negligible amount of mixing between $\mathcal{O}^S_{RR}$ and $\mathcal{O}^T_{RR}$ is produced by QED running~\cite{Datta:2020ocb}:
\beq
\begin{pmatrix}
	C^{V}_{RR}\\
	C^{S}_{RR}\\
	C^{T}_{RR}\\
\end{pmatrix}_{(m_b)} = \begin{pmatrix}
	1.0 & 0 & 0 \\
	0  & 1.2 & -1.5\times10^{-2}\\
	0  & -3.2\times 10^{-4} & 0.93 \\
\end{pmatrix} \begin{pmatrix}
	C^{V}_{RR}\\
	C^{S}_{RR}\\
	C^{T}_{RR}\\
\end{pmatrix}_{(M_Z)}\,.
\eeq

\subsection{Two-body $B$ meson decay in SMNEFT}
\label{sec:frameworkB}
In the SM, two-body $B$ decay to an electron and neutrino are helicity-suppressed. In SMNEFT, on the other hand, two-body decays can be enhanced by the RHN mass or the Wilson coefficients of helicity-flipped operators. The partial widths of the leptonic two-body decays $B^- \ra \ell^- \bar N$ through vector and scalar interactions are
\beqa
\Gamma_{B}(C^V_{RR}) &=& |C^V_{RR}|^2 \frac{f_B^2G_F^2 V_{cb}^2\sqrt{\lambda(m_B^2, m_N^2, m_{\ell}^2)}  }{8\pi} \frac{ m_B^2 (m_{\ell}^2+m_N^2)-(m_{\ell}^2-m_N^2)^2}{m_B^3}\,,\\
\Gamma_{B}(C^S_{RR,LR}) &=& |C^S_{RR,LR}|^2\frac{f_B^2G_F^2 V_{cb}^2\sqrt{\lambda(m_B^2, m_N^2, m_{\ell}^2)}  }{8\pi} \frac{ m_B  (m_B^2-m_{\ell}^2-m_N^2) }{(m_b+m_u)^2}\,,\label{eq:2decay}
\eeqa
where the decay constant $f_B = 0.19$ GeV~\cite{Zyla:2020zbs}. 
The triangle kinematic function $\lambda$ is defined as
\beq
\lambda(x,y,z) \equiv x^2+y^2+z^2 -  2xy - 2xz - 2yz\,.
\eeq
The branching fractions for $B^-\ra \ell^- \bar N$ are given in Table~\ref{Table: decay}. Since $B$ is a pseudoscalar meson, the tensor operator does not contribute to this decay.  For the helicity-flipped scalar operator $\mathcal{O}_{LNuQ}$, the decay width does not vanish in the massless limit of $\ell$ and $N$, as can be seen from Eq.~(\ref{eq:2decay}). In fact, the branching fractions do not depend strongly on $m_N$. Unlike the scalar interactions, the branching fractions for the vector operator $\mathcal{O}_{Nedu}$ are small for light $N$.
Another feature of two-body $B$ decays is that because the angular distribution of $\bar N$ in the $B$ rest frame is isotropic, it is independent of the $N$ production operator.

\begin{table}
	\centering
	\begin{tabular}{| c |  c |  c |  c | }
		\toprule
		three-body (two-body) & $\mathcal{O}_{LNuQ}$& $\mathcal{O}_{Nedu}$  & $\mathcal{O}^{(3)}_{LNQd}$   \\
		\midrule
		$m_N =$ 1 GeV  & $1.7~(14)$$|C_{LNuQ}|^2$ \%& $4.1~(0.4)$$|C_{Nedu}|^2$ \%& $40.5~(0)$ $|C^{(3)}_{LNQd}|^2$ \%\\
		\midrule
		$m_N =$ 100 MeV   &  $2.6~(15)$$|C_{LNuQ}|^2$ \%& $5.9~(4.3\times10^{-3})$ $|C_{Nedu}|^2$ \%& $59.3~(0)$$|C^{(3)}_{LNQd}|^2$  \%\\
		\midrule
		$m_N =$ 5 MeV   &  $2.6~(15)$$|C_{LNuQ}|^2$ \%& $5.9~(1.1\times10^{-5})$ $|C_{Nedu}|^2$ \%& $59.4~(0)$$|C^{(3)}_{LNQd}|^2$\%  \\
		\toprule
	\end{tabular}
	\caption{Branching fractions for  $\bar B^0 \ra D^{+} \ell^- \bar N,\,D^{*+} \ell^- \bar N$ and $B^-\ra \ell^- \bar N$ (in parenthesis).}
	\label{Table: decay}
\end{table}

\subsection{Three-body $B$ meson decay in SMNEFT}
\begin{figure}
	\centering
	\begin{subfigure}{.49\textwidth}
		\includegraphics[width=\textwidth]{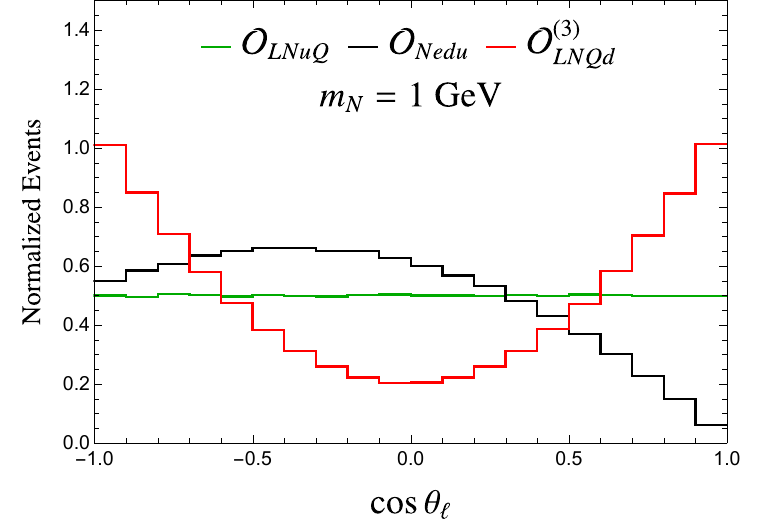}
		\label{fig:cos_1GeV}
	\end{subfigure}
	\begin{subfigure}{.49\textwidth}
		\includegraphics[width=\textwidth]{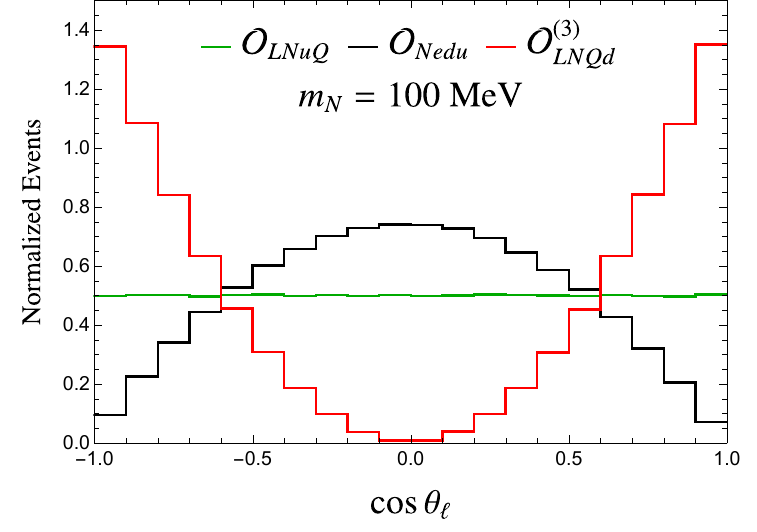}
		\label{fig:cos_100GeV}
	\end{subfigure}
	\caption{$\cos\theta_{\ell}$ distributions for $m_N= 1$~GeV (left panel) and 100~MeV (right panel) from $\bar B^0 \ra D^+ \ell^- \bar N$ with different production operators in SMNEFT. 
	Here $\theta_{\ell}$ is the angle between the charged lepton momentum in the $\ell \bar N$ rest frame and the $D$ momentum in the $\bar B$ rest frame.}
	\label{fig:Nellframe}
\end{figure}

The semileptonic three-body decays $\bar{B}^0 \ra D^{(*)+} \ell^- \bar N$ are described by the effective Lagrangian in Eq.~(\ref{eq:lag_B}) with $u$ replaced by $c$. 
The helicity amplitudes can be divided into four components according to the helicity combinations of $\ell$ ($\lambda_{\ell} =\pm 1/2$) and $\bar N$ ($\lambda_{\bar N} = \pm 1/2$). 
The differential decay widths are given by
\beqa
\frac{d^2\Gamma_{D}}{dq^2d\cos\theta_{\ell}} &=&\frac{1}{512 \pi^3} \frac{\lambda^{1/2}(m_B^2, m_{D}^2, q^2)}{m_B^3} \frac{\lambda^{1/2}(q^2,m_{\ell}^2, m_N^2)}{q^2}\sum_{\lambda_\ell,\lambda_{\bar N}}\overline{|M(\lambda_{\ell},\lambda_{\bar N})|^2},\label{eq:Ddecay} \\
\frac{d^4\Gamma_{D^*}}{dq^2d\cos\theta_{D}d\cos\theta_{\ell}d\varphi} &=&\frac{3}{2048 \pi^4}  \frac{\lambda^{1/2}(m_B^2, m_{D^*}^2, q^2)}{m_B^3} \frac{\lambda^{1/2}(q^2,m_{\ell}^2, m_N^2)}{q^2} \mathcal{B}(D^* \ra D\pi)\nonumber\\
&&\times \sum_{\lambda_\ell,\lambda_{\bar N}}\overline{|\sum_{\lambda_{D^*}}M(\lambda_{\ell},\lambda_{\bar N},\lambda_{D^*})|^2} \,.
\label{eq:Ddecay2}
\eeqa
where $q^2 \equiv (p_{\ell}+p_{N})^2$. Here, $\theta_{\ell}$ is the angle between the charged lepton momentum in the $\ell \bar N$ rest frame and the $D^{(*)}$ momentum in the $\bar B$ rest frame; the definitions of the other angles in Eq.~(\ref{eq:Ddecay2}), which we will not need here, can be found in Ref.~\cite{Datta:2022czw}. The helicity amplitudes of $\bar{B}^0 \ra D^{(*)+} \ell^- \bar N$ with nonzero $m_N$ have been calculated in Ref.~\cite{Datta:2022czw} and the hadronic form factors are given in Ref.~\cite{Bordone:2019guc}. 
The branching fractions for $\bar{B}^0 \ra D^{+} e^- \bar N, D^{*+} e^- \bar N$ are given in Table~\ref{Table: decay}.
Unlike the two-body $B$ meson decay, the $\cos \theta_\ell$ distributions depend on the $\bar N$ production operator. Figure~\ref{fig:Nellframe} shows the angular distributions for scalar $\mathcal{O}_{LNuQ}$, vector $\mathcal{O}_{Nedu}$, and tensor $\mathcal{O}^{(3)}_{LNQd}$ production operators with $m_N = 100$~MeV and $m_N = 1$ GeV. For $m_N < 100$~MeV, the angular distributions are similar to that of $m_N = 100$ MeV. The angular distributions are flat for the scalar interaction, and there is no forward-backward asymmetry for the tensor interaction. In the limit, $m_{\ell} \ra 0$, the forward-backward asymmetry for the vector operator increases with $m_N$~\cite{Datta:2022czw}.
The angular distributions of $\ell$ can be used to determine the Lorentz structure of the $\bar N$ production operators, as we will show below. 



\subsection{Polarization of $\bar N$ from $B$ decay }
In two-body $B$ decays, the angular distribution of $\bar N$ in the $B$ rest frame is isotropic. Any discrimination between different $\bar N$ production operators arises from the branching ratios and polarization of $\bar N$. 
A massive $\bar N$ from $B$ decay cannot be 100\% polarized. The $\bar N$  polarization affects the di-lepton angular distributions from three-body $\bar N$ decay significantly.
The polarization degree is defined as
\beq
P_{\bar N} = \frac{\Gamma_{\lambda_{\bar N} = -\frac{1}{2}}-\Gamma_{\lambda_{\bar N} = +\frac{1}{2}}}{\Gamma_{\lambda_{\bar N} = -\frac{1}{2}}+\Gamma_{\lambda_{\bar N} = +\frac{1}{2}}}\,,
\label{eq:P}
\eeq
where $\lambda_{\bar N}=+\frac{1}{2}$ ($-\frac{1}{2}$) denotes $\bar N$ having positive (negative) helicity. $\Gamma_{\lambda_{\bar N}= \pm \frac{1}{2}}$ is the $B$ meson decay width with $\lambda_{\bar N} = \pm \frac{1}{2}$. Equation~(\ref{eq:P}) is defined so that if $B$ decays to a massless left-handed antineutrino $\bar N$, $P_{\bar N} = 1$.

\begin{figure}
	\centering
	\includegraphics[width=0.5\columnwidth]{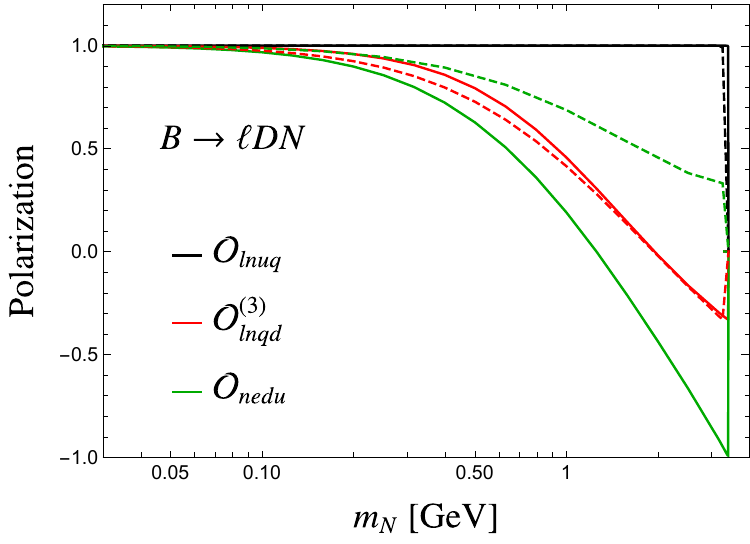}
	\caption{Fractional polarization of $\bar N$ as a function of heavy neutrino mass $m_N$. The black (red) [green] line corresponds to the case with $\bar{N}$ produced from a scalar ($\mathcal{O}_{LNuQ}$), tensor ($\mathcal{O}^{(3)}_{LNQd}$) or vector ($\mathcal{O}_{Nedu}$) operator. The solid (dashed) lines are for $\bar{B}^0 \ra D^{+}(D^{*+}) e^- \bar N$. 
	}
	\label{fig:pol}
\end{figure}

The polarization degree of $\bar N$ produced from $\bar{B}^0 \ra D^{(*)+} e^- \bar N$ via our three production operators is shown in Fig.~\ref{fig:pol}.
We see that the $\bar N$ produced from the scalar operator is 100\% polarized. The polarization is close to unity for $m_N \lesssim 100$~MeV for both decay channels and all three $ \bar N$ production operators. 
%
The values of $P_{\bar N}$ including both signal channels for $m_N= 1$~GeV, 100~MeV and 5~MeV are listed in Table~\ref{Table: pol}. For three-body $B$ decays, the $\bar N$ polarization produced by the scalar operator is close to unity below 2~GeV. In the tensor case, $P_{\bar N}$ can be as low as 0.42 (for $m_N= 1$~GeV), so that the angular distributions of the charged leptons from 
$\bar N$  decay are significantly averaged out.   
\begin{table}
	\centering
	\begin{tabular}{| c |  c |  c |  c | }
		\toprule
		three-body (two-body) & $\mathcal{O}_{LNuQ}$& $\mathcal{O}_{Nedu}$  & $\mathcal{O}^{(3)}_{LNQd}$   \\
		\midrule
		$m_N =$ 1 GeV  & 1.0 (1.0) & 0.49 (1.0)& 0.42 (-) \\
		\midrule
		$m_N =$ 100 MeV   & 1.0 (1.0) & 0.98 (1.0) & 0.98 (-)  \\
		\midrule
		$m_N =$ 5 MeV   & 1.0 (1.0) & 1.0 (1.0) & 1.0 (-)  \\
		\toprule
	\end{tabular}
	\caption{Polarization degree of $\bar N$ from $B$ meson decays.}
	\label{Table: pol}
\end{table}

For two-body $B$ decays via the scalar production operator $\mathcal{O}_{LNuQ}$, we find
\begin{eqnarray}
	P_{\bar N}(S)=\frac{\sqrt{1-{4y_N^2}/{(1-y_\ell^2+y_N^2)^2}}+\sqrt{1-{4y_\ell^2}/{(1+y_\ell^2-y_N^2)^2}}}{\sqrt{1-{4y_N^2}/{(1-y_\ell^2+y_N^2)^2}}\sqrt{1-{4y_\ell^2}/{(1+y_\ell^2-y_N^2)^2}}+1}\,,
\end{eqnarray}
where $y_{\ell} \equiv m_{\ell}/m_B$ and $y_{N} \equiv m_{N}/m_B$. Similarly, for the vector production operator $\mathcal{O}_{Nedu}$, 
\begin{eqnarray}
	P_{\bar N}(V)=\frac{\sqrt{1-{4y_N^2}/{(1-y_\ell^2+y_N^2)^2}}-\sqrt{1-{4y_\ell^2}/{(1+y_\ell^2-y_N^2)^2}}}{\sqrt{1-{4y_N^2}/{(1-y_\ell^2+y_N^2)^2}}\sqrt{1-{4y_\ell^2}/{(1+y_\ell^2-y_N^2)^2}}-1}\,,
\end{eqnarray}
which agrees with Eq.~(3.4) in Ref.~\cite{deGouvea:2021rpa}. 
As can be seen from Table~\ref{Table: pol}, $P_{\bar N}$ is always close to unity, except for the tensor operator which does not contribute to two-body $B$ decay.

\section{Three-body $\bar N$ decay}
\label{sec:FormalismN}

\subsection{$\bar N$ decay operators in the SMNEFT framework}
\label{sec:frameworkN}
Below the electroweak scale, the $\bar N$ decay $\bar N\ra e^+e^-\bar \nu $ can be described by four-fermion contact interactions. The effective Lagrangian of four-fermion operators above the electroweak scale in SMNEFT is
\beq
-\lag_{\text{SMNEFT}} \supset  2\sqrt{2} G_F [ C_{LN} \mathcal{O}_{LN} +  C_{Ne} \mathcal{O}_{Ne}+ C_{LNLe} \mathcal{O}_{LNLe} ]+ \text{h.c.}\,,
\eeq
where
\beqa
\mathcal{O}_{L N} &=& (\bar{N}\gamma_\mu N)(\bar{L}\gamma^\mu L)\,,\nonumber\\
\label{eq:Ndecay_opt}
\mathcal{O}_{Ne} &=& (\bar{N}\gamma_\mu N)(\bar{e}\gamma^\mu e)\,,\\
\mathcal{O}_{LNLe} &=& (\overline{L^j} N)\epsilon_{jk}(\overline{L^k} e)\,.\nonumber
\eeqa
Here, we keep the flavor indices implicit and assume $L$ and $e$ carry electron flavor only.

 At low energies, the general neutrino interaction (GNI) Lagrangian that describes $\bar N \ra e^+ e^-\bar\nu$ in the neutral current basis is, in the notation of Ref.~\cite{deGouvea:2021ual},
\beq
-\lag_{\text{GNI}} = \sum_{N,L}  (G_{NL} [\bar \nu \upgamma_N  N][\bar{e}\upgamma_L e] +\bar G_{NL} [\bar N \upgamma_N  \nu][\bar{e}\upgamma_L e] )+ \text{h.c.},
\eeq
where $\upgamma_N$, $\upgamma_L \in \{\mathbb{1},\,\gamma^5,\,\gamma^{\mu},\,\gamma^{\mu}\gamma^5,\sigma^{\mu\nu}\}$ are the scalar, pseudo-scalar, vector, axial-vector, and tensor  Lorentz structures, and $\nu$ denotes SM neutrinos or extra massless RHNs.
The nine independent $\bar G_{NL}$ dictating this decay are
\begin{equation}
	\left\lbrace \bar G_{SS}, \bar G_{SP}, \bar G_{PS}, \bar G_{PP}, \bar G_{VV}, \bar G_{VA}, \bar G_{AV}, \bar G_{AA}, \bar G_{TT} \right\rbrace.
\end{equation}
The matrix elements for both Majorana and Dirac $N$ are provided in Ref.~\cite{deGouvea:2021ual}.
We relate the SMNEFT operators to these nine operators by tree-level matching at the electroweak scale:
\beqa
\bar G_{SS} &=& -\bar  G_{SP} = -\bar  G_{PS} =\bar  G_{PP} = \frac{3}{8} C_{LNLe}\,,\\
\bar  G_{VV} &=&  \bar  G_{AV} = \frac{1}{4} (C_{Ne} + C_{LN} )\,,\\
\bar  G_{VA} &=& \bar  G_{AA} = \frac{1}{4} (C_{Ne} - C_{LN} )\,,\\
&\quad&\quad \bar  G_{TT} = \frac{1}{16}C_{LNLe}\,.
\eeqa
The mixing between the SMNEFT operators is given. by~\cite{Datta:2020ocb}
\beq
\begin{pmatrix}
	C_{LN}\\
	C_{Ne}\\
	C_{LNLe}\\
\end{pmatrix}_{(M_Z)} = \begin{pmatrix}
	1.0 & -1.3\times 10^{-3} & 0 \\
	-2.6\times 10^{-3}  & 1.0 & 0\\
	0  & 0 & 1.0 \\
\end{pmatrix} \begin{pmatrix}
	C_{LN}\\
	C_{Ne}\\
	C_{LNLe}\\
\end{pmatrix}_{(1~\text{TeV})}\,.
\eeq
and can be neglected.
Below the electroweak scale, there is no mixing between the neutral-current GNI operators.
Thus, we also neglect the effects of renormalization group running in the $\bar N$ decay operators. Although the three $\bar{N}$ production operators in Eq.~(\ref{eq:prod_opt}) permit decays to $\nu \gamma$ and $3\nu$ at loop level, the partial widths to these final states are negligible.

\subsection{Parametrization of three-body $\bar N$ decay }
\label{sec:3bodyN}
\begin{figure}[t]
	\centering
	\includegraphics[width=0.48\textwidth]{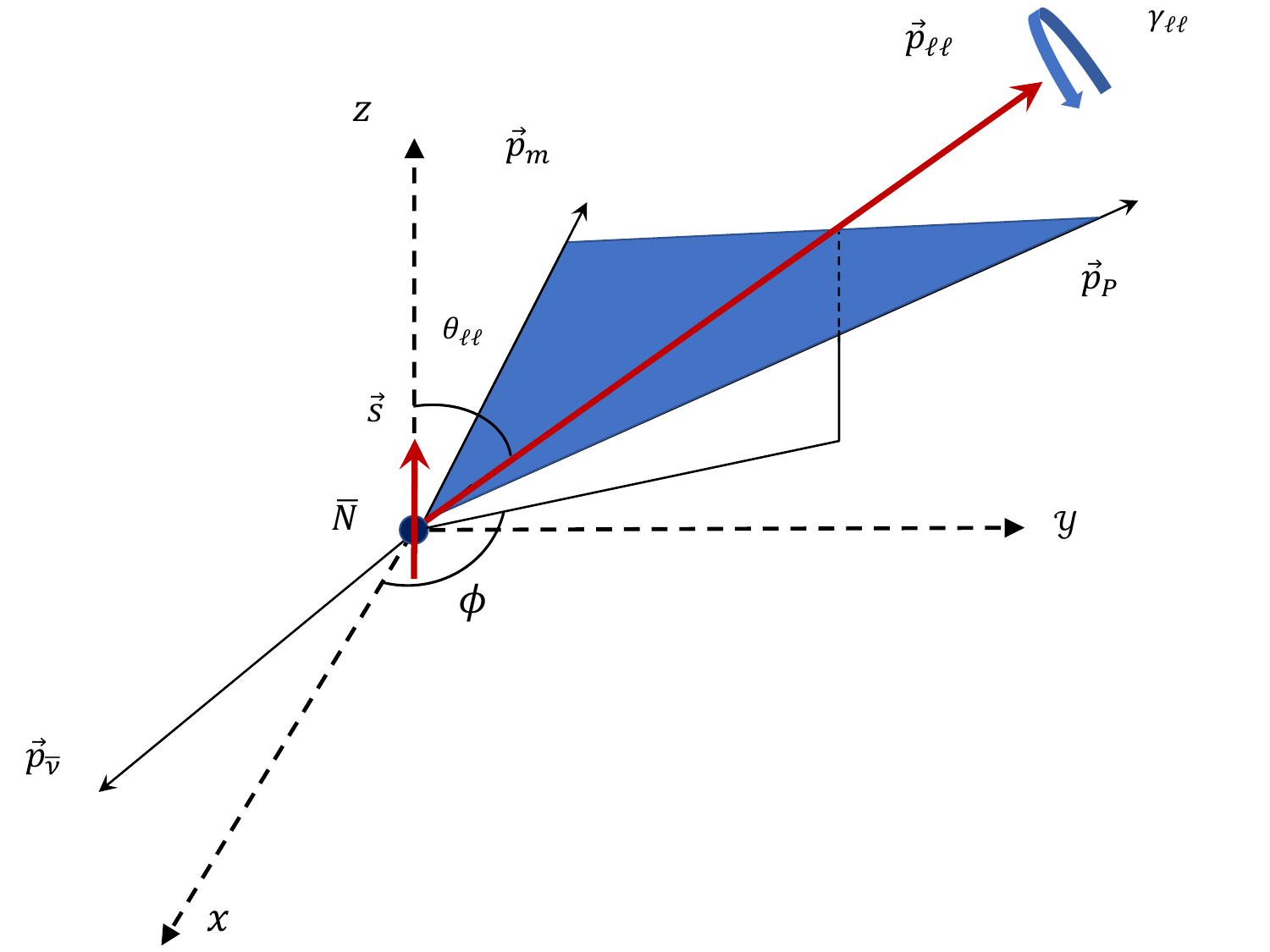}
	\includegraphics[width=0.48\textwidth]{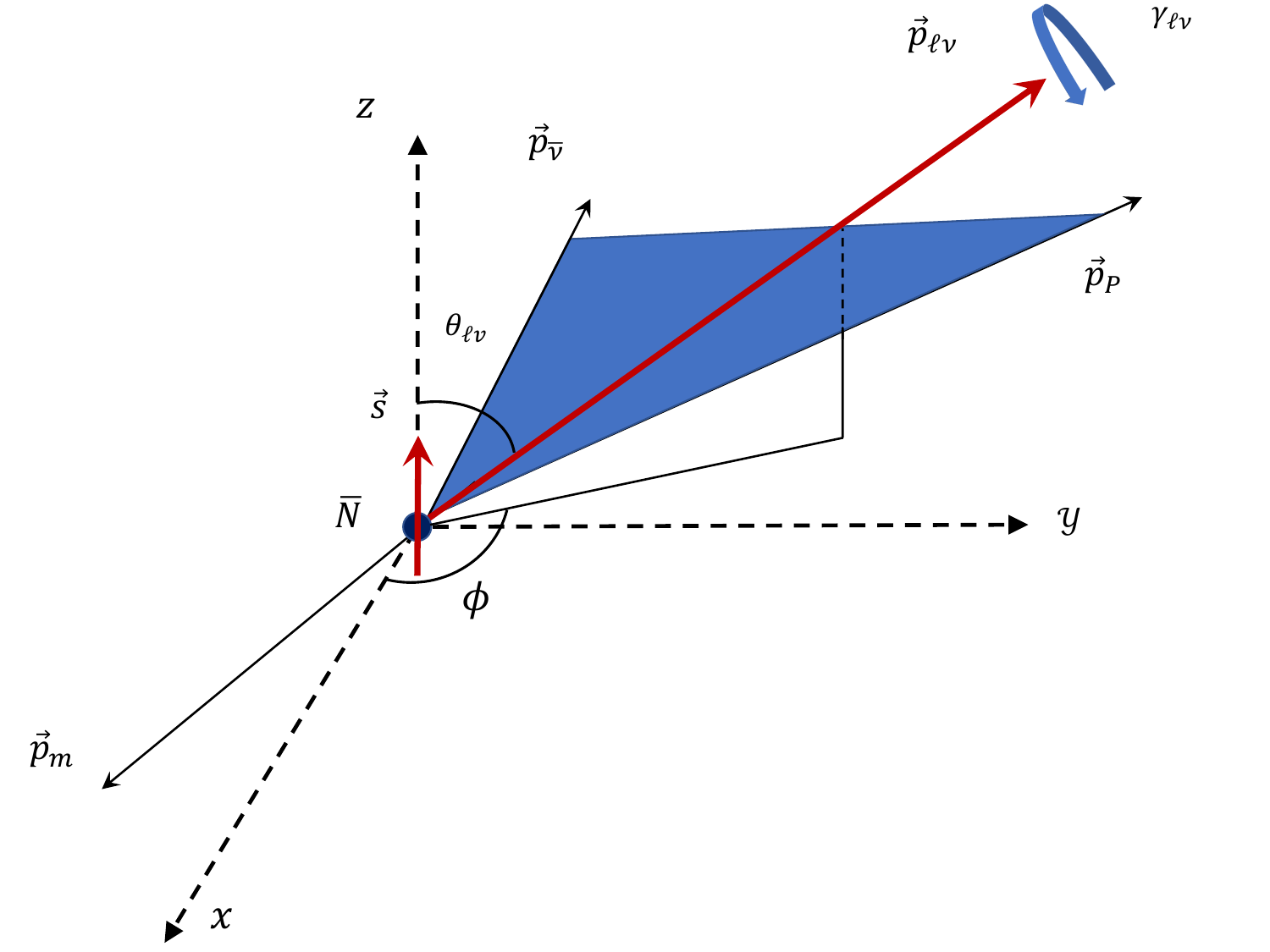}
	\caption{Kinematic variables for $\bar N \to \ell^- \ell^+\bar \nu$.}
	\label{fig:variables}
\end{figure}

In the $\bar N$ rest frame, $\bar N$ decay into the three-body final state $\ell^- \ell^+\bar \nu $ depends on five quantities:
\begin{equation}
	\left\lbrace m_{\ell\ell}^2,  m_{\nu m}^2, \cos\theta_{\ell\ell}, \gamma_{\ell\ell}, \phi\right\rbrace\,,
	\label{eq:ll}
\end{equation}
where $m_{\ell\ell}^2 \equiv \left(p_m^\mu + p_p^\mu\right)^2$ is the invariant mass-squared of the charged lepton pair with $p_m^\mu$ ($p_p^\mu$) the four-momentum of the negatively- (positively-)charged lepton, $m_{\nu m}^2 \equiv \left(p_m^\mu + p_\nu^\mu\right)^2$ is the invariant mass-squared of the neutrino/negatively-charged lepton system, $\theta_{\ell\ell}$ is the angle between the $\bar N$ spin direction and $\vec{p}_{\ell\ell} \equiv \vec{p}_{m}+\vec{p}_{p}$, 
$\gamma_{\ell\ell}$ is the angle of rotation of the charged-lepton subsystem about $\vec{p}_{\ell\ell}$, and $\phi$ is the azimuthal angle of $\vec{p}_{\ell\ell}$ about the spin direction; see the left panel of Fig.~\ref{fig:variables}. 
In terms of these quantities, the fully differential partial width for $\bar N$ decay can be written as~\cite{deGouvea:2021ual}
\begin{equation}\label{eq:diffxsec}
	\frac{d\Gamma\left(\bar N \to\ell^- \ell^+\bar  \nu \right)}{d \cos\theta_{\ell\ell} d\gamma_{\ell\ell} d m_{\ell\ell}^2 d m_{\nu m}^2 d\phi} = \frac{1}{\left(2\pi\right)^5} \frac{1}{64 m_N^3} \left\lvert \mathcal{M}(\bar N \to \ell^- \ell^+\bar \nu)\right\rvert^2,
\end{equation}
where $\left\lvert \mathcal{M}(\bar N \to\ell^- \ell^+\bar\nu)\right\rvert^2$ is the matrix-element-squared. 

\begin{figure}[tb]
	\centering
	\begin{subfigure}{.49\textwidth}
		\includegraphics[width=\textwidth]{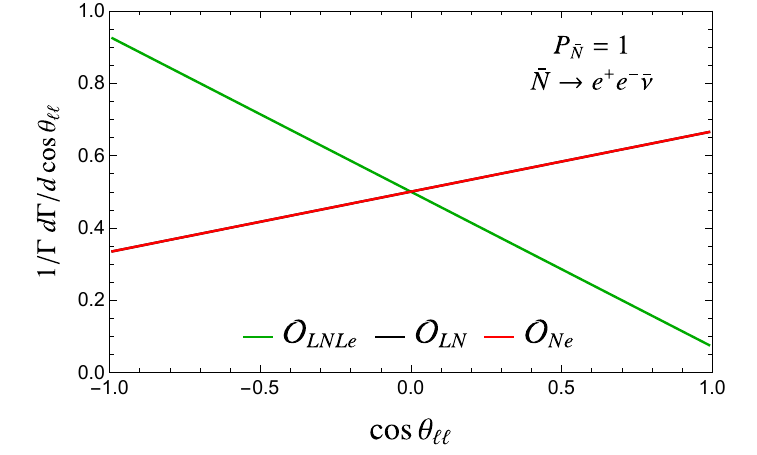}
		\label{fig:ll_cos_100_eee}
	\end{subfigure}
	\begin{subfigure}{.49\textwidth}
		\includegraphics[width=\textwidth]{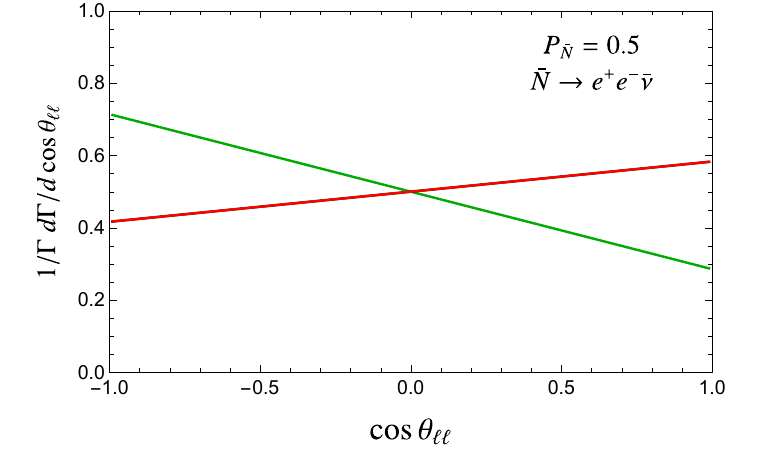}
		\label{fig:ll_cos_46_eee}
	\end{subfigure}
	\caption{$\cos \theta_{\ell\ell}$ distributions for $P_{\bar N} = 1$ (left panel) and $P_{\bar N} = 0.5$ (right panel), for three $\bar N$ decay operators: $\mathcal{O}_{LNLe}$ (green), $\mathcal{O}_{LN}$ (black), and  $\mathcal{O}_{Ne}$ (red). $N$ is assumed to be a DF. Note that the distributions for $\mathcal{O}_{LN}$ and $\mathcal{O}_{Ne}$ overlap completely.}
	\label{fig:cos_nosmear}
\end{figure}

After integration, we obtain the angular distributions in the $\bar N$ rest frame with respect to $\cos\theta_{\ell\ell}$ and $\gamma_{\ell\ell}$. If $N$ is a MF, the $\cos\theta_{\ell\ell}$ distributions are flat~\cite{BahaBalantekin:2018ppj}. In general, a large forward-backward asymmetry is possible for a DF.
In Fig.~\ref{fig:cos_nosmear}, we show the $\cos\theta_{\ell\ell}$ distributions for 100\% and 50\% polarized $\bar N$. The angular distributions are not sensitive to the RHN mass because detector effects are not yet included. A larger polarization makes it easier to distinguish between a Dirac neutrino and a Majorana neutrino. In fact, a Dirac neutrino with a vanishing polarization behaves like a Majorana neutrino in the decay. 

\begin{figure}[tb]
	\centering
	\begin{subfigure}{.32\textwidth}
		\includegraphics[width=\textwidth]{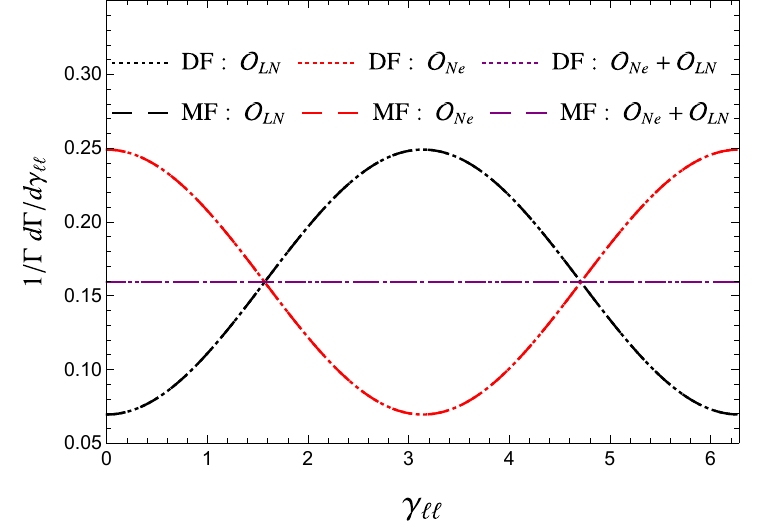}
	\end{subfigure}
	\begin{subfigure}{.32\textwidth}
	\includegraphics[width=\textwidth]{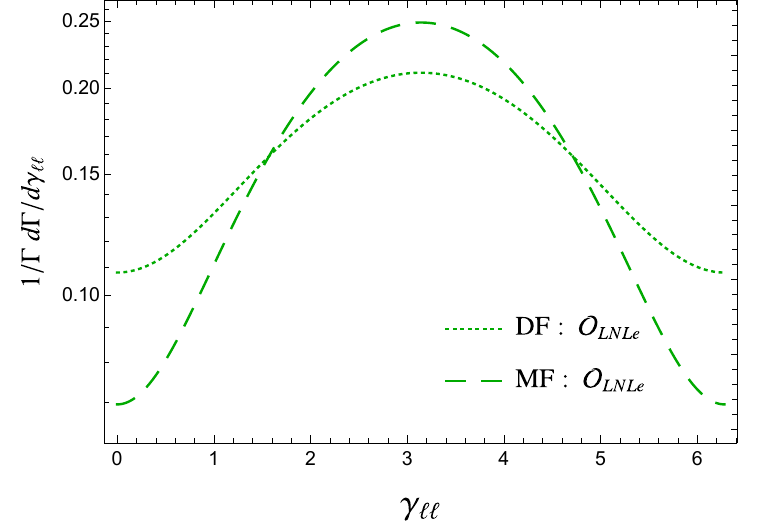}
	\end{subfigure}
	\begin{subfigure}{.32\textwidth}
		\includegraphics[width=\textwidth]{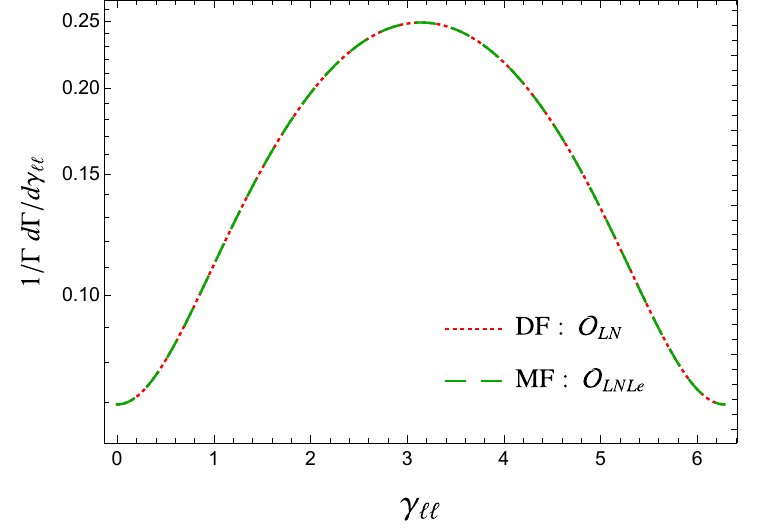}
	\end{subfigure}
	\caption{$\gamma_{\ell\ell}$ distributions with $P_{\bar N} = 1$ in the $\bar N$ rest frame. Left panel: the distributions for two vector $\bar N$ decay operators: $\mathcal{O}_{LN}$~(black), $\mathcal{O}_{Ne}$~(red) for DF~(dotted lines) and MF (dashed lines) decays. The DF and MF distributions are identical. Middle panel: the distributions for the scalar operator $\mathcal{O}_{LNLe}$ for DF~(dotted lines) and MF (dashed lines) decays. Right panel: the distributions for DF and vector operator $\mathcal{O}_{LN}$~(dotted red), and MF and scalar operator $\mathcal{O}_{LNLe}$~(dashed green), overlap completely. }
	\label{fig:ga_nosmear}
\end{figure}

 The $\gamma_{\ell\ell}$ distributions from a MF decay are not necessarily flat and while they may be sensitive to the $\bar N$ decay operators, they may not be useful to distinguish DF from MF. For example, as shown in the left panel of Fig.~\ref{fig:ga_nosmear}, DF and MF have the same $\gamma_{\ell\ell}$ distributions for the vector operators $\mathcal{O}_{LN}$ and $\mathcal{O}_{Ne}$;
 note that the linear combination of these operators yields flat $\gamma_{\ell\ell}$  distributions. 
 The distributions for the scalar operator $\mathcal{O}_{LNLe}$ are different for DF and MF as can be seen from the middle panel. It is also interesting that $\mathcal{O}_{LN}$ for DF can mimic the $\gamma_{\ell\ell}$ distribution of $\mathcal{O}_{LNLe}$ for MF as shown in the right panel of Fig.~\ref{fig:ga_nosmear}. In this case, we are not able to tell the nature of neutrinos or the interaction type from only the $\gamma_{\ell\ell}$ distribution. 

\section{Bounds on the Wilson coefficients}
\label{sec:Bounds}

Stringent bounds on the WCs of the production operators can be obtained from tests of lepton flavor universality  in $\pi$ and $K$ decays~\cite{Bryman:2019bjg}. However, in the case of flavor non-universal scenarios, these bounds are highly dependent on the flavors associated with these operators. Denote $C^{\alpha\beta\gamma\delta}$ as the WC of a production operator with flavor indices $\alpha$, $\beta$, $\gamma$ and $\delta$ for the heavy neutrinos, charged leptons, up-type quarks and down-type quarks, respectively. Since most lepton flavor universality constraints depend on both $C^{\alpha1\gamma\delta}$ and $C^{\alpha2\gamma\delta}$, if both are allowed to be nonzero, no bounds can be imposed on the WCs with either charged lepton flavor~\cite{Han:2020pff}. Also, if $N$ is not kinematically accessible, the measured branching factions in heavy meson decays impose stringent constraints on the WCs of the production operators~\cite{Bryman:2019bjg}. To evade these constraints, we assume that all the WCs are zero except those associated with the electrons and bottom quarks, i.e., only $C^{\alpha1\gamma3}$ are nonzero. Then, bounds from the branching fractions of heavy meson decays, such as those listed in Table 3 of Ref.~\cite{Atre:2009rg}, do not apply.

Since there is no experimental constraint on $B^+ \to e^+e^-e^+ \nu_e$ for prompt $\bar N$ decays,
%
we place bounds on the $\bar N$ production operators using the uncertainties in the branching fractions of the inclusive decay modes $B^+ \to \ell^+ \nu_{\ell}+\text{anything}$ and $B^0 \to \ell^+ \nu_{\ell}+\text{anything}$, which is 0.28\%~\cite{Zyla:2020zbs}. 
The $1\sigma$ upper bounds obtained by requiring $\mathcal{B}(B^- \to \ell^- \bar N ) < 0.28$\% and $\mathcal{B}(\bar B^0 \to D^{+} \ell^- \bar N,  D^{*+} \ell^- \bar N ) < 0.28$\%, are shown in Table~\ref{Table: bound}.
\begin{table}
	\centering
	\begin{tabular}{| c |  c |  c |  c | }
		\toprule
		three-body (two-body) & $C_{LNuQ}$& $C_{Nedu}$  & $C^{(3)}_{LNQd}$   \\
		\midrule
		$m_N =$ 1 GeV  & $0.41~(0.14)$  & $0.26~(0.84)$ & $0.083~(-)$ \\
		\midrule
		$m_N =$ 100 MeV   & $0.33~(0.14)$  & $0.22~(
		1.0)$ & $0.069~(-)$   \\
		\midrule
		$m_N =$ 5 MeV   & $0.33~(0.14)$  & $0.22~(
		1.0)$ & $0.069~(-)$    \\
		\toprule
	\end{tabular}
	\caption{$1\sigma$ upper bounds on the $\bar N$ production operators from measurements of the inclusive decays, $B \ra \ell + \text{anything}$.}
	\label{Table: bound}
\end{table}
We do not take any WC to be larger than unity. Note that the use of branching fraction uncertainties only provides a rough estimate.  A more accurate determination of existing limits requires a recast of the analyses in Refs.~\cite{Belle:2015pkj, Belle:2017rcc}, which is beyond the scope of this work.

 %

\section{Results at Belle II}
\label{sec:Belle}
The Belle II experiment~\cite{Belle-II:2018jsg} 
will perform a wide range of high-precision measurements of the products of $B$ meson decays. About 50 billion $\Upsilon(4S)$ resonances will be collected at Belle II with an integrated luminosity of 50~ab$^{-1}$. The decay branching ratio of $\Upsilon(4S)$ to $B^+B^-$ is 51.4\% and to $B^0\bar{B}^0$ is 48.6\%~\cite{Zyla:2020zbs}.

\subsection{Belle II simulation}

\subsubsection{$\bar N$ signal reconstruction}

In our simulation, we rely on the fully tagged $B$ meson decay via charged tracks on one side ($B_\text{tag}$).
For the signal, we include the leptonic two-body decay $B^-\ra \ell^- \bar N$ and semi-leptonic three-body decay $\bar{B}^0 \ra D^{(*)+} \ell^- \bar N$  channels. 
Then, the $\bar N$ momentum ($p_{\bar N}$) can be fully reconstructed kinematically:
\beq
p_{e^+} + p_{e^-} = p_{\rm tag}  + p_{D^{(*)+} \ell^-} + p_{\bar N}\,,
\label{eq:pn}
\eeq
and the signal mass peak can be obtained by the recoil mass technique without relying on the $\bar N$ decay products,
\beq
m_N^2 = (p_{e^+} + p_{e^-} - p_{\rm tag}  - p_{D^{(*)+} \ell^-})^2\,.
\label{eq:mn}
\eeq

\begin{figure}
	\centering
	\includegraphics[width=0.5\columnwidth]{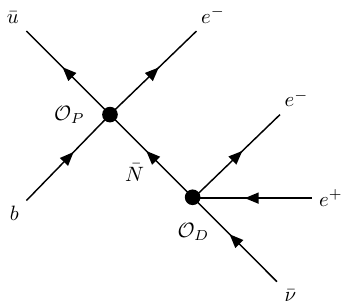}
	\caption{Two-body $B$ meson decay chain. $\mathcal{O}_{P}$ is one of the three $\bar N$ production operators in Eq.~(\ref{eq:prod_opt}) and $\mathcal{O}_{D}$ is one of the three $\bar N$ decay operators in Eq.~(\ref{eq:Ndecay_opt}). The three-body $B$ decay is the same but with $\bar u$ replaced by $\bar c$.
	}
	\label{fig:Bdecay}
\end{figure}

The decay chain of the $B$ meson and $\bar N$ is depicted in Fig.~\ref{fig:Bdecay}. Since there is one missing neutrino in the well-constrained kinematics,  the four-momentum of the light antineutrinos in the subsequent decay $\bar N\ra e^+e^- \bar \nu$
can be determined on an event-by-event basis. 

\subsubsection{Simulation details}

\begin{figure}
        \centering
 	\includegraphics[width=0.4\columnwidth]{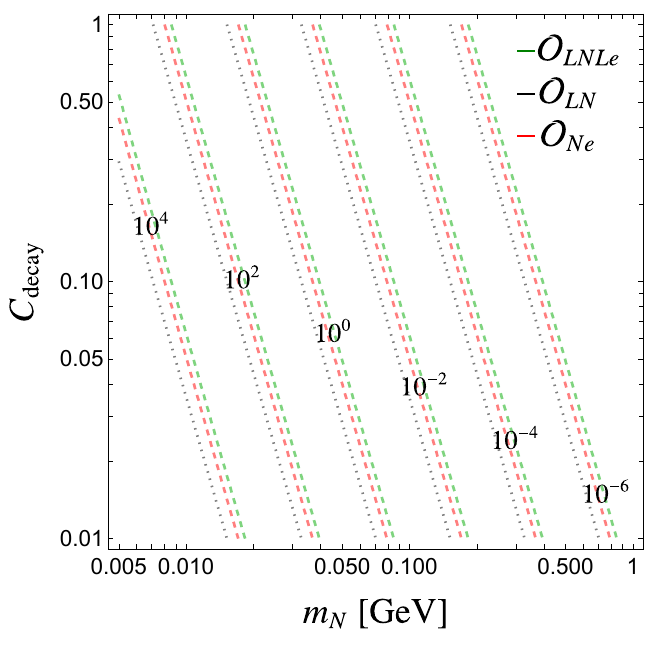}
	 \includegraphics[width=0.4\columnwidth]{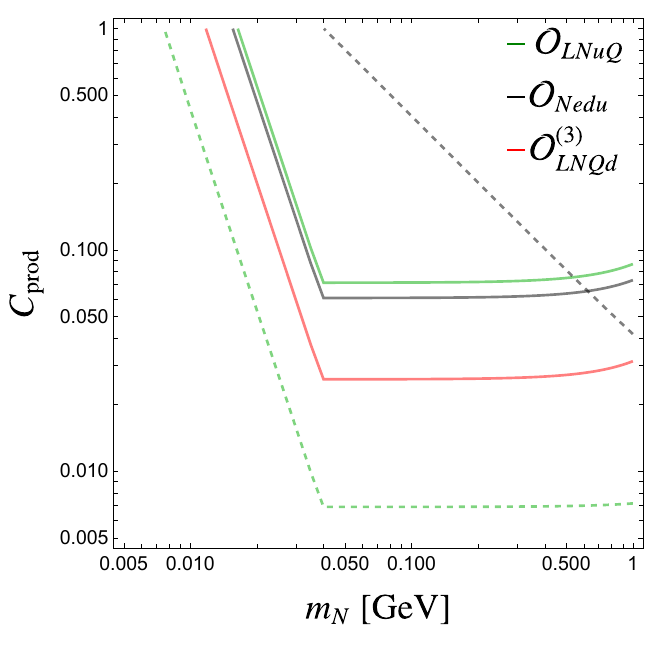}
 	\caption{Left panel: Iso-decay length contours  in the lab frame (in meters)  for the three decay operators, $\mathcal{O}_{LNLe}$ (green), $\mathcal{O}_{LN}$ (black), and $\mathcal{O}_{Ne}$ (red). Right panel: $5\sigma$ discovery sensitivity for the three production operators, $\mathcal{O}_{LNuQ}$ (green), $\mathcal{O}_{Nedu}$ (black), and $\mathcal{O}^{(3)}_{LNQd}$ (red). The solid (dashed) lines are for three-body (two-body) $B$ decay. 
 	}
 	\label{fig:decay}
 \end{figure}

 We take into account the realistic energy, transverse momentum and angular smearing of charged particles using~\cite{Adachi:2018qme, Expt}
 \beqa
 &&\sigma_{p_T}/p_T = 0.0011 p_T~\text{[GeV]} \oplus 0.0025/\beta\,,\nonumber\\
 &&\sigma_{E}/E = \text{2.25\% (1 GeV) and 7.7\% (0.1 GeV)}\,,\nonumber\\
 &&\sigma_\theta (m_N = 100, 1000~\text{MeV}) =6~\text{mrad}\,, \ \ \ \sigma_\phi (m_N = 100,1000~\text{MeV})  =9~\text{mrad}\,,\nonumber\\
 &&\sigma_\theta (m_N = 5~\text{MeV}) =\sigma_\phi (m_N = 5~\text{MeV})  =5~\text{mrad}\,.
 \label{eq:smear}
 \eeqa
 Then we apply the following cuts on the transverse momentum $p^e_{T}$ and polar angle $\theta_{e}$:
 \beq
 p_T^{e} > 200~\text{MeV }\text{and } 17^{\circ} <\theta_{e}< 150^{\circ}\,.
 \label{eq:cut1}
 \eeq
 From the positron and two electrons in the final state, we need to identify the $e^+e^-$ pair from $\bar N$ decay and reconstruct the $\cos\theta_{\ell\ell}$ and $\gamma_{\ell\ell}$ angular distributions. To select the correct pair, we make use of the opening angle between the electron and positron and the reconstructed RHN mass. For $m_N \leq 100$~MeV, we select the pair with the smaller opening angle. For $m_N = 1$~GeV, we identify the electron from the $B$ decay as the one that gives a reconstructed $m_N$ (in Eq.~\ref{eq:mn}) closer to 1~GeV. To remove the SM background we also apply a cut on $m_{ee}$ following the results of a detector simulation~\cite{Expt},
 \beq
 m_{ee} < m_N + 3\Gamma_N\,,
 \label{eq:cut2}
 \eeq
 where $\Gamma_N$ is the $\bar N$ decay width. The detector simulation shows that the SM background is negligible and the $\bar N$ reconstruction efficiency is not sensitive to the value of $\Gamma_N$ for promptly decaying $\bar N$~\cite{Expt}; if $\Gamma_N$ is too small, most $\bar N$ will decay outside the Belle~II detector. 
 Neglecting effects of final state radiation, we find the probability of correct pairing to be larger than 98\%.
 The decay length in the lab frame assuming only one decay channel $\bar N \to e^+e^-\bar\nu$ is shown in the left panel of Fig.~\ref{fig:decay}. 
 
 For $m_N > 100$~MeV, it is safe to assume that $\bar{N}$ decays via the three decay operators with $\mathcal{O}(1)$ WCs. The corresponding $\mathcal{B}(\bar N\to e^+e^-\bar\nu) = 0.5$ for $\mathcal{O}_{LN}$, and unity for 
 $\mathcal{O}_{LNLe}$ and $\mathcal{O}_{Ne}$. On the other hand, for $m_N = 5$~MeV, the WCs need to be of $\mathcal{O}(10^2)$ to ensure that $\bar N$ decays inside the Belle~II detector. In order to have a 5~MeV RHN with decay length $\sim 10^{-2}$~m in the lab frame, we take $\mathcal{B}(\bar N\to e^+e^-\bar\nu) = 10^{-5}$ and $\mathcal{O}(1)$ WCs for the $\bar N$ decay operators. For three-body decay, we include events from both $\bar{B}^0 \ra D^{+}e^-\bar N$ and $\bar{B}^0 \ra D^{*+}e^-\bar N$, but select signal events only from $D^+ \to \pi^+\pi^+K^-$ and $D^{*+} \to (D^0 \to K^-\pi^+)\pi^+$, which have a 9.4\% and 2.6\% branching fraction, respectively.
 The hadronic tag-side efficiency for $B^0\bar{B}^0$ and $B^+B^-$ is 0.18\% and 0.28\%, respectively~\cite{Keck:2018lcd}.
 Based on these assumptions, the signal events in three-body and two-body $B$ meson decays are given by
 \beqa
 N_{\text{2-body}}^{\text{sig}} &=& N_{\Upsilon} \cdot \mathcal{B}(\Upsilon\to B^+B^-) \cdot \epsilon_{\text{tag}} (B^+) \cdot  \mathcal{B}(B^- \to e^- \bar N) \cdot \mathcal{B}(\bar N\to e^+e^-\bar\nu) \cdot \epsilon_{\text{rec}} (\bar N)\cdot \epsilon_{\text{cut}} \,,
 \nonumber\\
 N_{\text{3-body}}^{\text{sig}}(D) &=& N_{\Upsilon} \cdot \mathcal{B}(\Upsilon\to B^0\bar B^0) \cdot \epsilon_{\text{tag}} (B^0) \cdot \mathcal{B}(\bar{B}^0 \to D^{+} e^- \bar N) \cdot  \mathcal{B}(D^+\to K^-\pi^+\pi^+) 
 \nonumber\\
 &\,\,&\times \mathcal{B}(\bar N\to e^+e^-\bar\nu) \cdot \epsilon_{\text{rec}} (\bar N)\cdot \epsilon_{\text{cut}} \,,\nonumber\\
  N_{\text{3-body}}^{\text{sig}} (D^*)&=& N_{\Upsilon} \cdot \mathcal{B}(\Upsilon\to B^0\bar B^0) \cdot \epsilon_{\text{tag}} (B^0) \cdot \mathcal{B}(\bar{B}^0 \to D^{*+} e^- \bar N) \cdot \mathcal{B}(D^{*+} \ra D^0 \pi^+) 
 \nonumber\\
 &\,\,&\times \mathcal{B}( D^0 \ra K^-\pi^+)\cdot \mathcal{B}(\bar N\to e^+e^-\bar\nu) \cdot \epsilon_{\text{rec}} (\bar N)\cdot \epsilon_{\text{cut}} \,,
 \eeqa
 where $N_{\Upsilon}=50\times 10^9$, the reconstruction efficiency $\epsilon_{\text{rec}}(\bar N)=0.05$. We assume that the total branching fraction of the two decay channels is 
 $\mathcal{B}(\bar{B}^0 \to D^{+} e^- \bar N\,, D^{*+} e^- \bar N) = 0.28\%$ to obtain the maximum number of signal events consistent with current bounds. The $\bar B^0\to D^{*+}e^-\bar N: \bar B^0\to D^+e^-\bar N$ signal event ratio for $m_N = 1.0$~GeV (5~MeV) is 0.080~(0.093), 1.56~(1.66), and 31.4~(38.2) for the production operators, $\mathcal{O}_{LNuQ}$, $\mathcal{O}_{Nedu}$, and $\mathcal{O}^{(3)}_{LNQd}$, respectively. The signal event ratio for $m_N =100$~MeV is essentially the same as for  $m_N =5$~MeV. Correspondingly,
 for $m_N = 1.0$~GeV~(5~MeV and 100~MeV), $\mathcal{B}(\bar{B}^0 \to D^{+} e^- \bar N) = 0.259~(0.256)\%,\,0.110~(0.105)\%$ and 0.00863~(0.00714)\% for the production operators, $\mathcal{O}_{LNuQ}$, $\mathcal{O}_{Nedu}$, and $\mathcal{O}^{(3)}_{LNQd}$, respectively. 
  
 The efficiencies of the cuts in Eqs.~(\ref{eq:cut1}) and (\ref{eq:cut2}) $\epsilon_{\text{cut}}$ are typically 40 - 60\% for $m_N = 1$~GeV and 20 - 30\% for $m_N \lesssim 100$~MeV. 
 The maximum number of expected signal events for either $\mathcal{O}_{Ne}$ and $\mathcal{O}_{LNLe}$ are given in Table~\ref{Table: event}. For the decay operator $\mathcal{O}_{LN}$, the maximum number of expected signal events is halved because decay to a three-neutrino final state is also available. 
Note that $D^+\to K^-\pi^+\pi^+\pi^0$, $D^{*+} \ra (D^0 \ra K^-\pi^+\pi^0) \pi^+$ and $D^{*+} \ra (D^+ \ra K^-\pi^+\pi^+) \pi^0$ with an unidentified $\pi^0$ introduces an extra effective smearing in the signal event distributions.  The $\pi^0$ identification efficiency is 98\%~(93\%) if the threshold for both photons is 20~(30) MeV, values that are attainable in the future~\cite{Belle-II:2018jsg}. 
The corresponding contamination in $\bar{B}^0 \ra D^{+}e^-\bar N$ and $\bar{B}^0 \ra D^{*+}e^-\bar N$ is 1\%~(5\%)  and 9\%~(33\%), respectively.
In our analysis we take the $\pi^0$ efficiency to be 98\% and neglect the extra smearing due to an unidentified $\pi^0$.

The parameter region in which a $5\sigma$ observation can be attained with 25 signal events is shown in the right panel of Fig.~\ref{fig:decay}. 
Note that for $m_N=5$~MeV, it will not be possible to obtain enough events at Belle II with 50 ab$^{-1}$.
Nevertheless, we do not discard this scenario, as it might be relevant with a higher integrated luminosity and improvements in efficiencies. 

\begin{table}
	\centering
	\begin{tabular}{| c |  c |  c |  c | }
		\toprule
		three-body (two-body) & $\mathcal{O}_{LNuQ}$& $\mathcal{O}_{Nedu}$  & $\mathcal{O}^{(3)}_{LNQd}$   \\
		\midrule
		$m_N =$ 1 GeV  & $2.5~(62)\times 10^{2}$  & $1.4~(61)\times 10^{2}$ & $0.76~(-)\times 10^{2}$ \\
		\midrule
		$m_N =$ 100 MeV   & $0.97~(31)\times 10^{2}$ & $0.57~(0.47)\times 10^{2}$  & $0.31~(-)\times 10^{2}$  \\
		\midrule
		$m_N =$ 5 MeV   &$1.1~(34)\times 10^{-3}$ & $0.57~(1.3\times10^{-3})\times 10^{-3}$ & $0.31~(-)\times 10^{-3}$   \\
		\toprule
	\end{tabular}
	\caption{The maximum number of signal events for either of the decay operators, $\mathcal{O}_{Ne}$ and $\mathcal{O}_{LNLe}$, expected at Belle II with 50~ab$^{-1}$. }
	\label{Table: event}
\end{table}

   \begin{figure}[tb]
 	\centering
 	\begin{subfigure}{.32\textwidth}
 		\includegraphics[width=\textwidth]{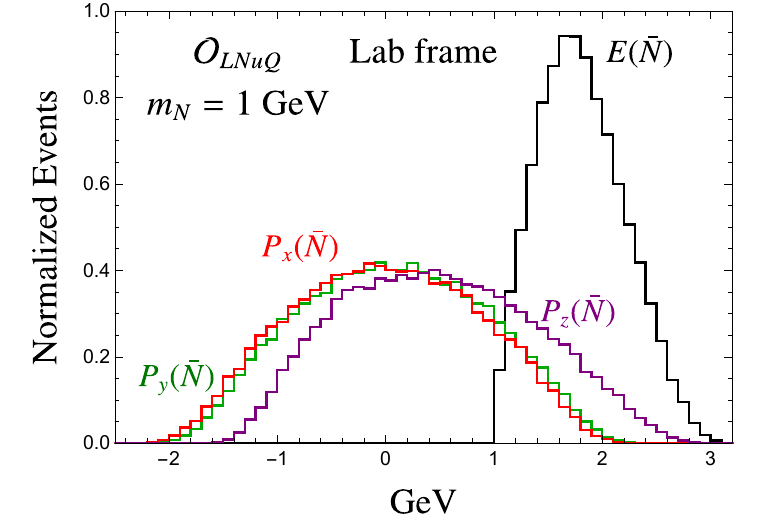}
 		\label{fig:N_4p_lnuq_Lab_1GeV}
 	\end{subfigure}
 	\begin{subfigure}{.32\textwidth}
 		\includegraphics[width=\textwidth]{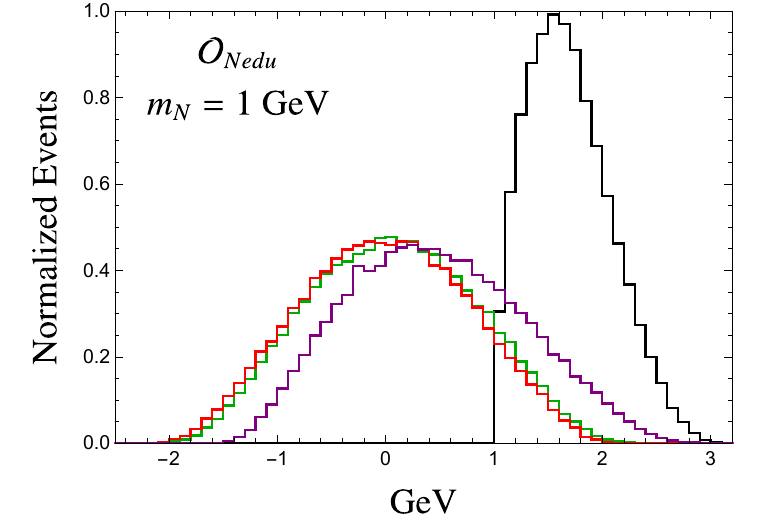}
 		\label{fig:N_4p_nedu_Lab_1GeV}
 	\end{subfigure}
 	\begin{subfigure}{.32\textwidth}
 		\includegraphics[width=\textwidth]{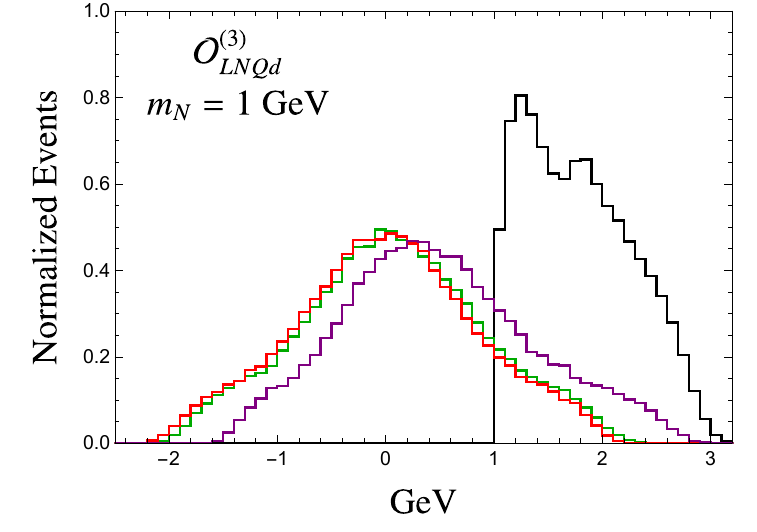}
 		\label{fig:N_4p_lnqd3_Lab_1GeV}
 	\end{subfigure}
 	\begin{subfigure}{.32\textwidth}
 		\includegraphics[width=\textwidth]{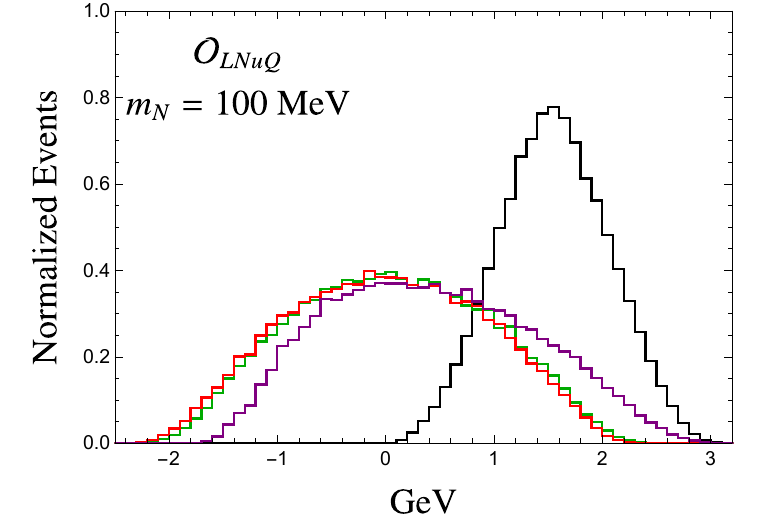}
 		\label{fig:N_4p_lnuq_Lab_100MeV}
 	\end{subfigure}
 	\begin{subfigure}{.32\textwidth}
 		\includegraphics[width=\textwidth]{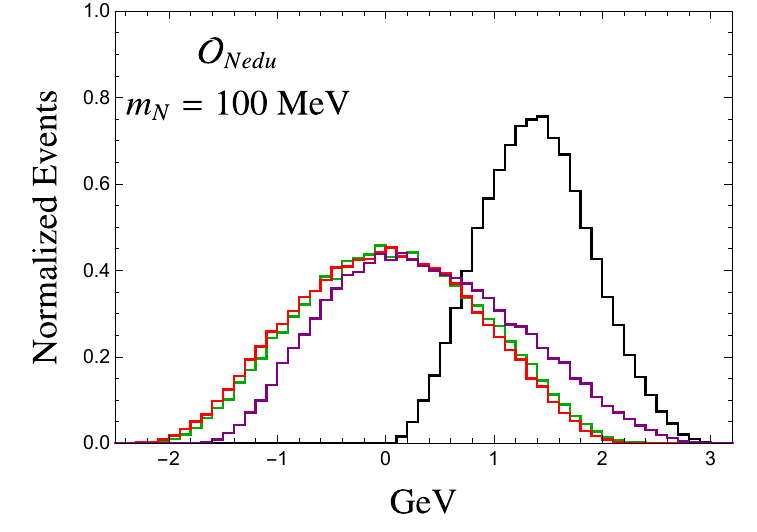}
 		\label{fig:N_4p_nedu_Lab_100MeV}
 	\end{subfigure}
 	\begin{subfigure}{.32\textwidth}
 		\includegraphics[width=\textwidth]{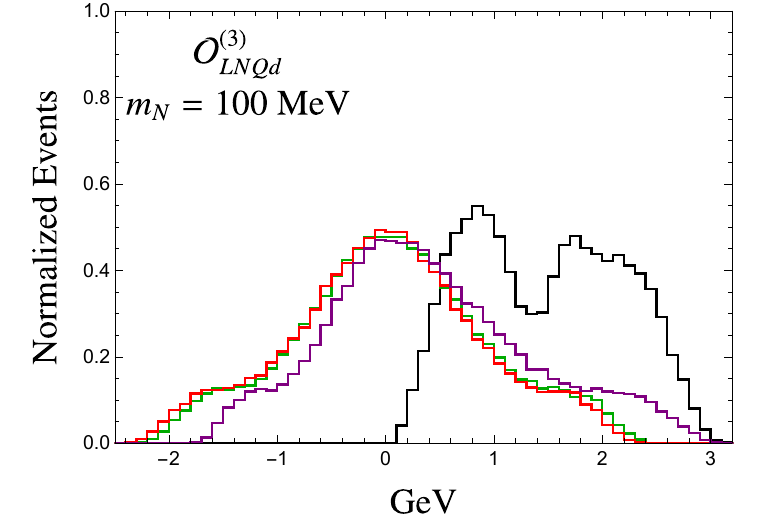}
 		\label{fig:N_4p_lnqd3_Lab_100MeV}
 	\end{subfigure}
 	\caption{Four-momentum distributions for $\bar N$ in the lab frame from three-body $B$ decay for the three production operators with $m_N = 1$~GeV (upper panels) and $m_N=100$~MeV (lower panels). The four-momentum distributions for $m_N=5$~MeV are similar to those for  $m_N=100$~MeV.}
 	\label{fig:4momentum_Lab}
 \end{figure}

 The details of our simulations of the angular distributions for the two-body and three-body decays of $B$ mesons at Belle~II are as follows:
 \begin{enumerate}
 	\item We use the $B$ meson four-momentum distributions from simulations based on the fully reconstructed $B_\text{tag}$~\cite{Expt}.
 	\item For the decays, $\bar{B}^0 \ra D^{+} \ell^- \bar N$ and $B^- \ra \ell^- \bar N$, we simulate $\bar N$ four-momentum distributions in the $\ell \bar N$ rest frame according to Eq.~(\ref{eq:Ddecay}) with the three $\bar N$ production operators in SMNEFT: $\mathcal{O}_{LNuQ}$, $\mathcal{O}_{Nedu}$, and $\mathcal{O}^{(3)}_{LNQd}$. The $\cos \theta_\ell$ distributions are shown in Fig.~\ref{fig:Nellframe}. 
 	\item We boost $\bar N$ to the lab frame by using the $B$ meson four-momentum in step 1. 
 	The four-momentum distributions of $\bar N$ in the lab frame are shown in Fig.~\ref{fig:4momentum_Lab}.
 	\item For  $\bar N\ra \ell^+\ell^-\bar \nu$, we generate the four-momentum distributions of $\ell^+\ell^-$ in the $\bar N$ rest frame according to the formulas given in Ref.~\cite{deGouvea:2021ual} with the three $\bar N$ decay operators in SMNEFT: $\mathcal{O}_{LNLe}$, $\mathcal{O}_{LN}$, $\mathcal{O}_{Ne}$. The angular distributions are shown in Figs.~\ref{fig:cos_nosmear} and~\ref{fig:ga_nosmear}.
 	\item We then boost $\ell^+\ell^-$ to the lab frame by using the four-momentum of $\bar N$ in the lab frame from step 3.
 	\item We smear the energy and momentum of the charged particles and apply cuts on the transverse momentum $p^{\ell}_T$, polar angule $\theta_\ell$, and di-lepton invariant mass $m_{ee}$ according to Eqs.~(\ref{eq:smear})-(\ref{eq:cut2}).
 	\item Finally, we boost the smeared charged lepton four-momentum to the $\bar N$ rest frame and extract the values of $\cos\theta_{\ell\ell}$ and $\gamma_{\ell\ell}$. 
	The smeared angular distributions are shown in Fig.~\ref{fig:angle_2body} for two-body $B$ decay and Fig.~\ref{fig:angle_3body} for three-body $B$ decay.
 \end{enumerate}

 \subsection{Determining the Dirac or Majorana nature of $N$}
 \label{sec:results2b}
 We now study the potential of Belle~II to distinguish the Dirac versus Majorana nature of heavy neutrinos by employing the 
 $\cos\theta_{\ell\ell (\nu)}$ and $\gamma_{\ell\ell (\nu)}$ angular distributions. Since these are defined in the $\bar N$ rest frame (as shown in Fig.~\ref{fig:variables}), we combine the $\bar{B}^0 \ra D^{+}e^-\bar N$ and $\bar{B}^0 \ra D^{*+}e^-\bar N$ channels in our analysis. We neglect any differences in smearing in the two channels.

 \subsubsection{Results from two-body $B$ decay }
\begin{figure}[tb]
 	\centering
 \begin{subfigure}{.32\textwidth}
	\includegraphics[width=\textwidth]{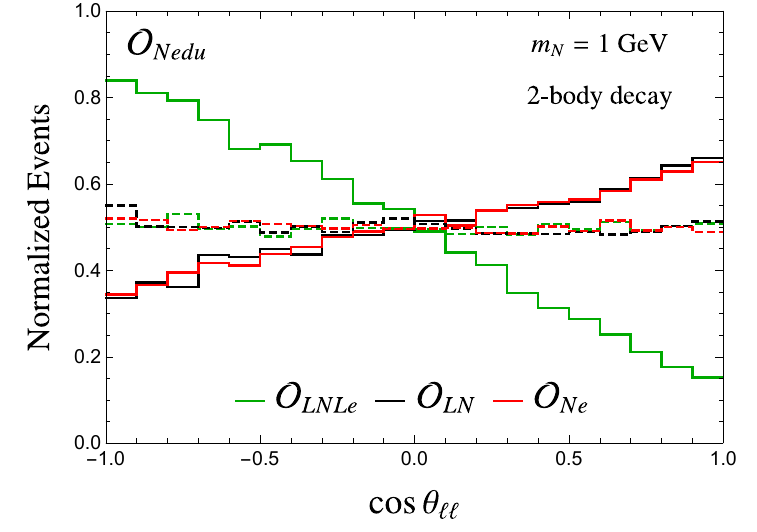}
	\label{fig:cut_smear_ll_cos_nedu_1GeV_eee_2body}
\end{subfigure}
\begin{subfigure}{.32\textwidth}
	\includegraphics[width=\textwidth]{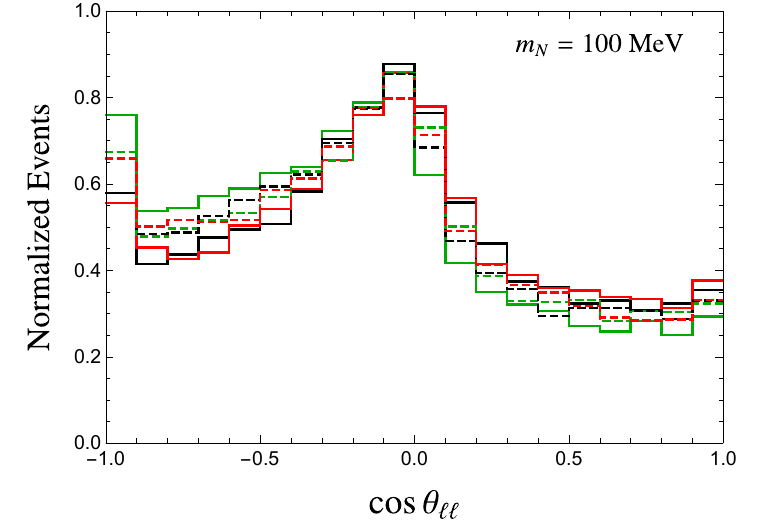}
	\label{fig:cut_smear_ll_cos_nedu_100MeV_eee_2body}
\end{subfigure}
\begin{subfigure}{.32\textwidth}
	\includegraphics[width=\textwidth]{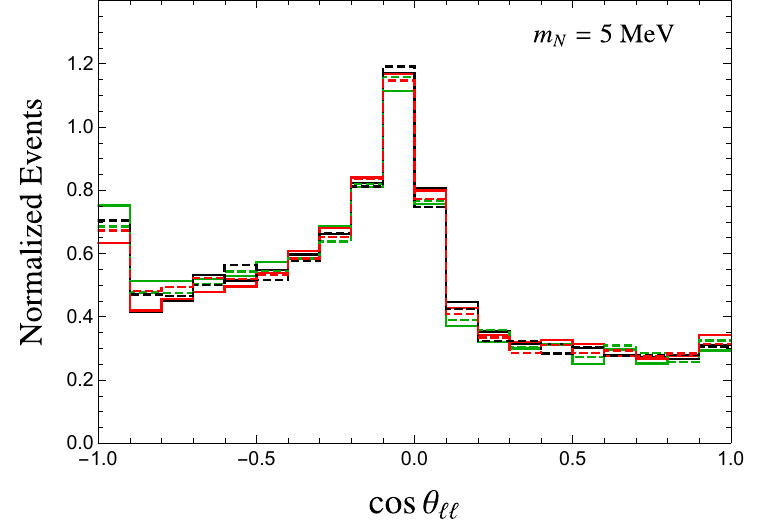}
	\label{fig:cut_smear_ll_cos_nedu_5MeV_eee_2body}
\end{subfigure}
	\begin{subfigure}{.32\textwidth}
	\includegraphics[width=\textwidth]{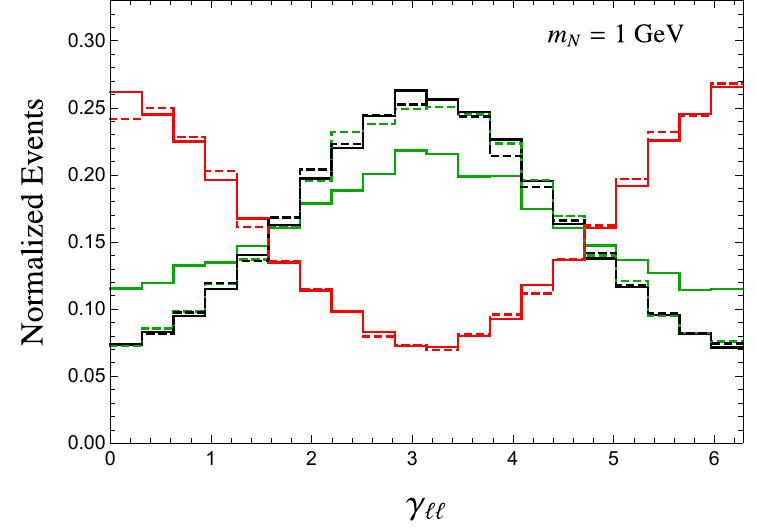}
	\label{fig:cut_smear_ll_ga_nedu_1GeV_eee_2body}
\end{subfigure}
\begin{subfigure}{.32\textwidth}
	\includegraphics[width=\textwidth]{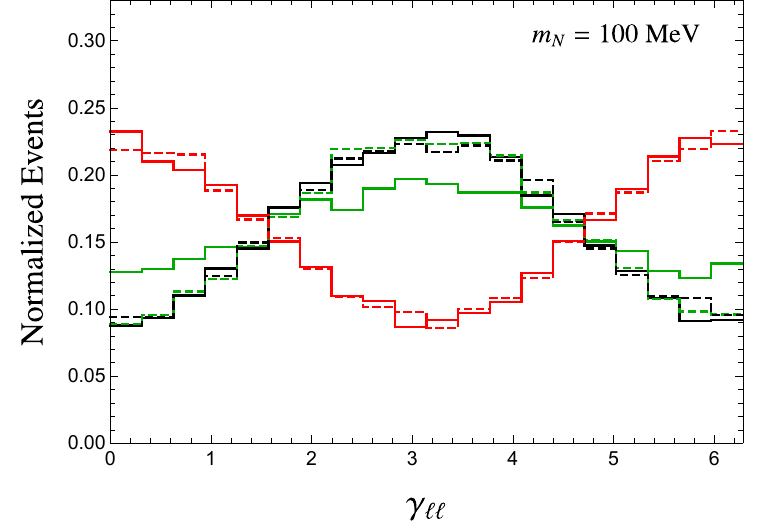}
	\label{fig:cut_smear_ll_ga_nedu_100MeV_eee_2body}
\end{subfigure}
\begin{subfigure}{.32\textwidth}
	\includegraphics[width=\textwidth]{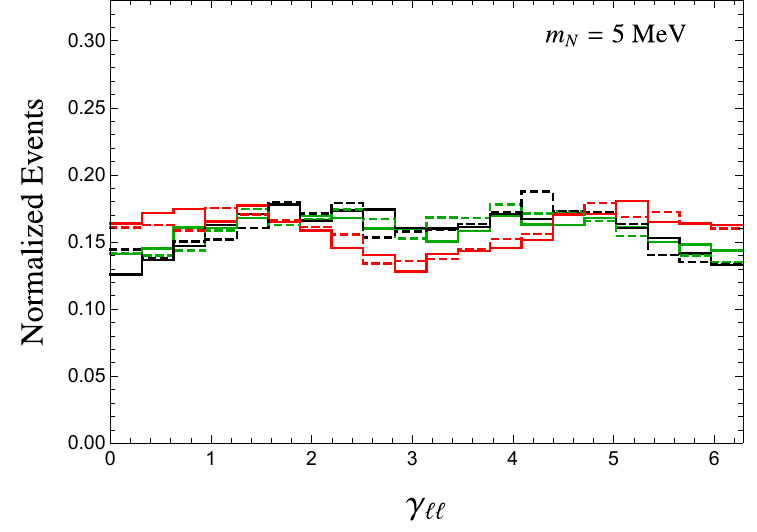}
	\label{fig:cut_smear_ll_ga_nedu_5MeV_eee_2body}
\end{subfigure}
 	\caption{Smeared $\cos \theta_{\ell\ell}$ distributions (upper panels) and $\gamma_{\ell\ell}$ distributions (lower panels) for the production operator $\mathcal{O}_{Nedu}$ in the $\bar N$ rest frame from two-body $B$ decay for the three decay operators: $\mathcal{O}_{LNLe}$ (green), $\mathcal{O}_{LN}$ (black), $\mathcal{O}_{Ne}$ (red). Solid (dashed) lines are for a  DF (MF) $\bar N$.}
 	\label{fig:angle_2body}
 \end{figure}

 We first consider the case with $\bar N$ produced in two-body leptonic $B$ decay. 
 The $\cos \theta_{\ell\ell}$ and $\gamma_{\ell\ell}$ angular distributions including detector effects in the $\bar N$ rest frame for the production operator $\mathcal{O}_{Nedu}$ and the three decay operators, $\mathcal{O}_{LNLe}$, $\mathcal{O}_{LN}$ and $\mathcal{O}_{Ne}$, for DF (solid lines) and MF (dashed lines)  are shown in Fig.~\ref{fig:angle_2body}. 
The angular distributions of the production operator $O_{LNuQ}$ are very similar to those of $O_{Nedu}$, which implies that the angular distributions  from two-body $B$ decay are not sensitive to the SMNEFT production operators.  Also, the two-body decay process is not sensitive to the production operator $\mathcal{O}^{(3)}_{LNQd}$ due to the pseudoscalar nature of the $B$ meson.
 
 For $m_N$ =1~GeV, the $\cos\theta_{\ell\ell}$ distributions are not sensitive to detector effects as the $\bar N$ is not highly boosted and the angular distance between the di-leptons is large. However, for \mbox{$m_N \lesssim 100$~MeV}, the ability to utilize the information in the $\cos\theta_{\ell\ell}$ distributions is very limited as the $\bar N$ is highly boosted, and the di-leptons are squeezed together. The $\gamma_{\ell\ell}$  distribution is less sensitive to smearing effects. The spectral features are unspoiled by detector effects for $m_N = 100~$MeV, but are smeared out for $m_N = 5$~MeV. It becomes more difficult to resolve the nature of heavy neutrinos as $m_N$ becomes smaller.
 The $\gamma_{\ell\ell}$ distributions for DF and MF are the same except for $\mathcal{O}_{LNLe}$. 
It is easier to distinguish Dirac from Majorana neutrinos using the $\cos \theta_{\ell\ell}$ distribution than the $\gamma_{\ell\ell}$ distribution. 

\begin{figure}[tb]
	\centering
	\begin{subfigure}{.32\textwidth}
		\includegraphics[width=\textwidth]{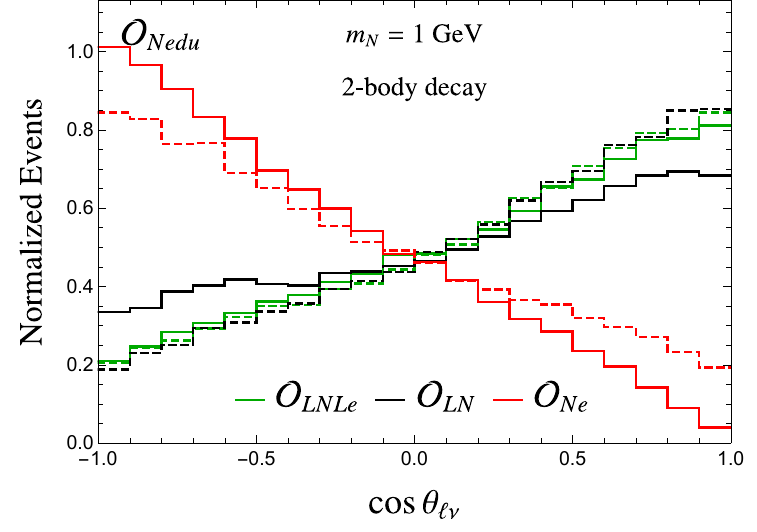}
		\label{fig:cut_smear_ln_cos_nedu_1GeV_eee_2body}
	\end{subfigure}
	\begin{subfigure}{.32\textwidth}
		\includegraphics[width=\textwidth]{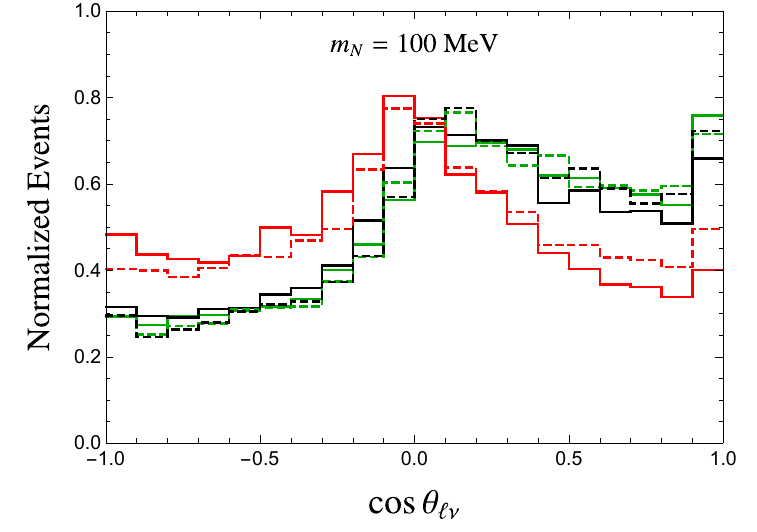}
		\label{fig:cut_smear_ln_cos_nedu_100MeV_eee_2body}
	\end{subfigure}
	\begin{subfigure}{.32\textwidth}
		\includegraphics[width=\textwidth]{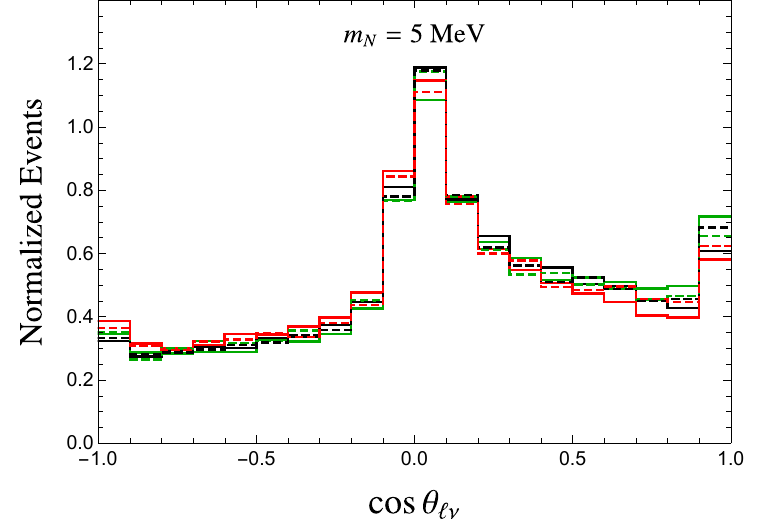}
		\label{fig:cut_smear_ln_cos_nedu_5MeV_eee_2body}
	\end{subfigure}
	\begin{subfigure}{.32\textwidth}
		\includegraphics[width=\textwidth]{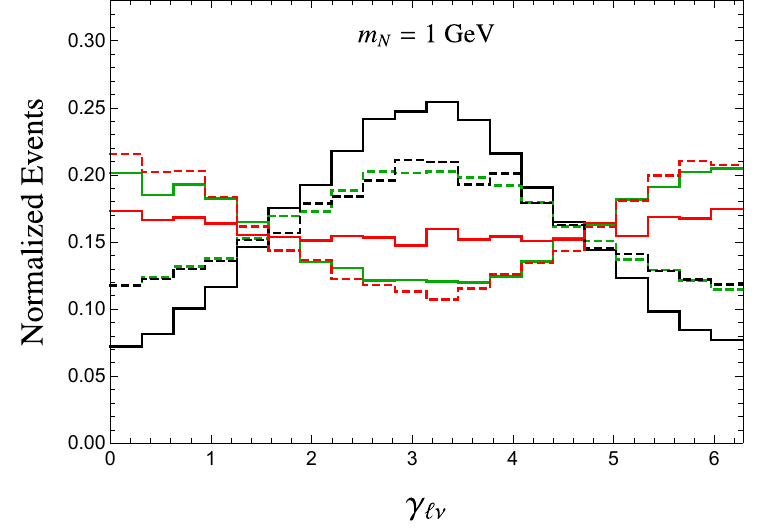}
		\label{fig:cut_smear_ln_ga_nedu_1GeV_eee_2body}
	\end{subfigure}
	\begin{subfigure}{.32\textwidth}
		\includegraphics[width=\textwidth]{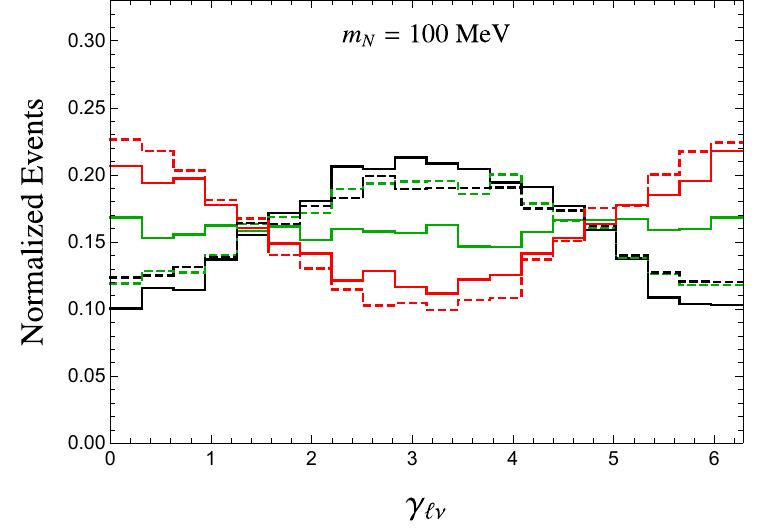}
		\label{fig:cut_smear_ln_ga_nedu_100MeV_eee_2body}
	\end{subfigure}
	\begin{subfigure}{.32\textwidth}
		\includegraphics[width=\textwidth]{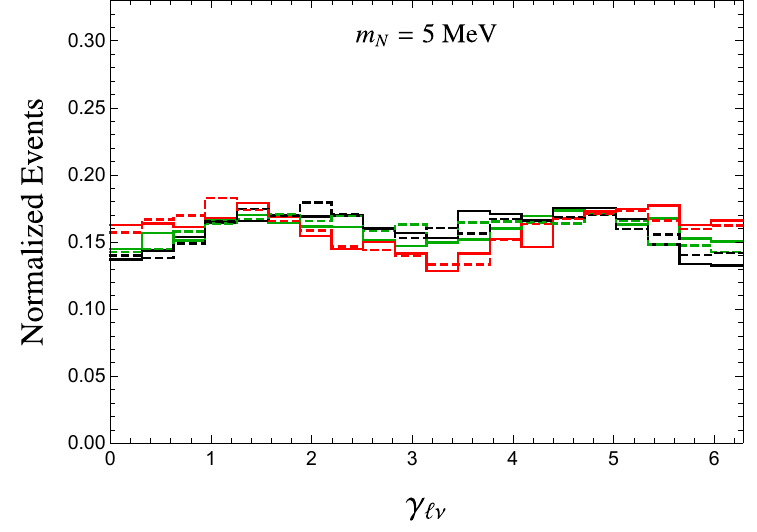}
		\label{fig:cut_smear_ln_ga_nedu_5MeV_eee_2body}
	\end{subfigure}
	\caption{Same as Fig.~\ref{fig:angle_2body}, except for angular observables associated with $\ell\nu$. }
	\label{fig:angle_2body_ln}
\end{figure}

Since the $\cos \theta_{\ell\ell}$ distribution is sensitive to smearing effects, we define the angular observables $\cos\theta_{\ell\nu}$ and $\gamma_{\ell\nu}$, whose definitions are the same as $\cos\theta_{\ell\ell}$ and 
$\gamma_{\ell\ell}$ but with $\ell^-$ replaced by $\bar \nu$; see the right panel of Fig.~\ref{fig:variables}. Their distributions from two-body $B$ decay are shown in Fig.~\ref{fig:angle_2body_ln}. We find that $\gamma_{\ell\nu}$ can be useful to distinguish the nature of $N$, and $\cos\theta_{\ell\nu}$ can be useful to distinguish the $\bar N$ decay operators if $N$ is a MF.

To evaluate the statistical significance with which a Dirac $N$ can be distinguished from a Majorana $N$, we define a $\chi^2$ function suited to a Poisson distribution of events,
\begin{align}
\chi^2 = \sum_{i=1}^{\rm N_{bin}}  2\left[ (1+\alpha_i)N_{i}^{\rm th} - N_{i}^{\rm obs} + N_{i}^{\rm obs}\ln\frac{N_{i}^{\rm obs}}{ (1+\alpha_i)N_{i}^{\rm th}}\right]+\frac{\alpha_i^2}{\sigma_s^2}\,,
\label{eq:chi2}
\end{align}
where $N_{i}^{\rm obs}$  and $N_{i}^{\rm th}$ are the event counts per bin for a Dirac or a Majorana  fermion, respectively. We take the total number of events to be 100  and the number of bins $N_{\rm bin}=20$. We also introduce 20 nuisance parameters $\alpha_i$ for the per-bin systematic scalings with penalty $\sigma_s=10\%$, which accounts for hadronic form factor uncertainties and statistical fluctuations in our simulation. We marginalize over $\alpha_i$ to obtain the minimum $\chi^2$, which serves an estimate of the ability to distinguish the Dirac versus Majorana nature of $N$,  i.e., we use $\chi^2_{\rm min}=9.0\  (25.0)$ as our $3\sigma$ ($5\sigma$) criterion.
 The values of $\chi^2_{\rm min}$ for two-body $B$ decay are listed in Table~\ref{Table: 2body}.  We see that the sensitivities for the production operators  $\mathcal{O}_{LNuQ}$ and
  $\mathcal{O}_{Nedu}$ are about the same. For the decay operator $\mathcal{O}_{ LNLe}$, the $\cos \theta_{\ell\ell}$ distribution provides the most sensitive probe to the nature of RHNs, while for the decay operators $\mathcal{O}_{LN}$ and $\mathcal{O}_{Ne}$, the $\cos \theta_{\ell\nu}$ distribution is the most sensitive probe. Also, with 100~events,  the  nature of $N$ can be determined at more than $5\sigma$~CL for all operator combinations in Table~\ref{Table: 2body} for $m_N = 1$~GeV, but not for $m_N = 100$~MeV and 5~MeV. 

\begin{table}
	\centering
	\renewcommand{\arraystretch}{1.5}
	\tabcolsep=0.1cm
	\begin{tabular}{|c|c|c|c|c|c|c|c|c|c|c|c|c|}
		\toprule
		Production & \multicolumn{12}{|c|}{$\mathcal{O}_{LNuQ}$} \\
		\hline
		Decay & \multicolumn{4}{|c|}{$\mathcal{O}_{LNLe}$} & \multicolumn{4}{|c|}{$\mathcal{O}_{LN}$}& \multicolumn{4}{|c|}{$\mathcal{O}_{Ne}$} \\
		\hline
		Distribution & $\cos \theta_{\ell\ell}$ & $\gamma_{\ell\ell}$ & $\cos \theta_{\ell\nu}$ &$\gamma_{\ell\nu}$ 
		& $\cos \theta_{\ell\ell}$ & $\gamma_{\ell\ell}$ & $\cos \theta_{\ell\nu}$ &$\gamma_{\ell\nu}$ 
		& $\cos \theta_{\ell\ell}$ & $\gamma_{\ell\ell}$ & $\cos \theta_{\ell\nu}$ &$\gamma_{\ell\nu}$ \\
		\hline
		$m_N = 1$ GeV&142&12&1.2&43  &25&0.5&27&13  &24&0.3&46&9.5        \\
		\midrule
		$m_N = 100$ MeV  &9.7&5.6&1.5&9.4  &4.7&0.3&4.1&2.5  &6.6&0.7&8.4&2.1        \\
		\midrule
		$m_N = 5$ MeV  &1.6&1.2&1.7&1.7  &1.1&0.7&1.1&0.3  &2.2&0.5&1.4&0.4       \\
		
		\toprule
		Production & \multicolumn{12}{|c|}{$\mathcal{O}_{Nedu}$} \\
		\hline
		Decay & \multicolumn{4}{|c|}{$\mathcal{O}_{ LNLe}$} & \multicolumn{4}{|c|}{$\mathcal{O}_{LN}$}& \multicolumn{4}{|c|}{$\mathcal{O}_{Ne}$} \\
		\hline
		Distribution & $\cos \theta_{\ell\ell}$ & $\gamma_{\ell\ell}$ & $\cos \theta_{\ell\nu}$ &$\gamma_{\ell\nu}$ 
		& $\cos \theta_{\ell\ell}$ & $\gamma_{\ell\ell}$ & $\cos \theta_{\ell\nu}$ &$\gamma_{\ell\nu}$ 
		& $\cos \theta_{\ell\ell}$ & $\gamma_{\ell\ell}$ & $\cos \theta_{\ell\nu}$ &$\gamma_{\ell\nu}$ \\
		\hline
		$m_N = 1$ GeV&137&11&1.3&41  &25&0.2&31&12  &23&0.3&44&10        \\
		\midrule
		$m_N = 100$ MeV  &7.5&8.1&2.2&12 &7.8&0.6&5.6&2.3  &7.1&0.4&7.7&2.1          \\
		\midrule
		$m_N = 5$ MeV  &2.4&0.8&2.6&0.8  &2.4&1.0&1.4&0.5  &2.4&0.6&2.4&0.5      \\
		\toprule
	\end{tabular}
	\caption{{\it Distinguishing between Dirac and Majorana $N$}. $\chi^2_{\rm min}$ for  two-body $B$ decay for various production and decay operators in SMNEFT. }
	\label{Table: 2body}
\end{table}

 \begin{figure}[tb]
	\centering
	\begin{subfigure}{.32\textwidth}
		\includegraphics[width=\textwidth]{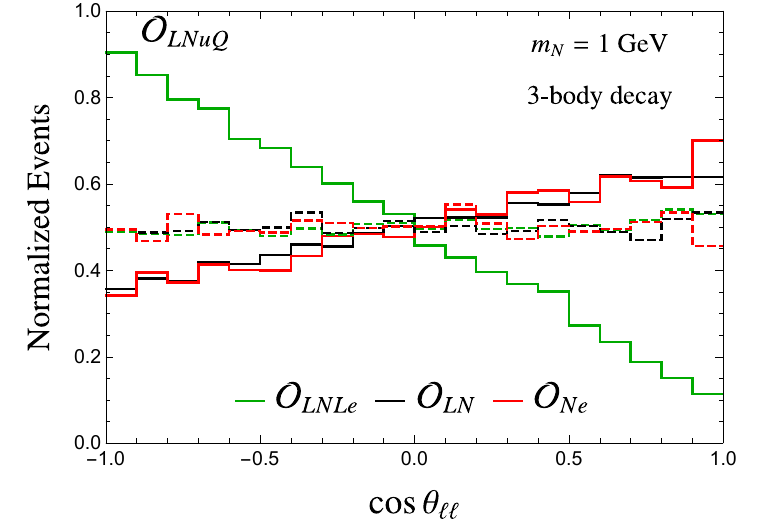}
		\label{fig:cut_smear_ll_cos_lnuq_1GeV_eee}
	\end{subfigure}
	\begin{subfigure}{.32\textwidth}
	\includegraphics[width=\textwidth]{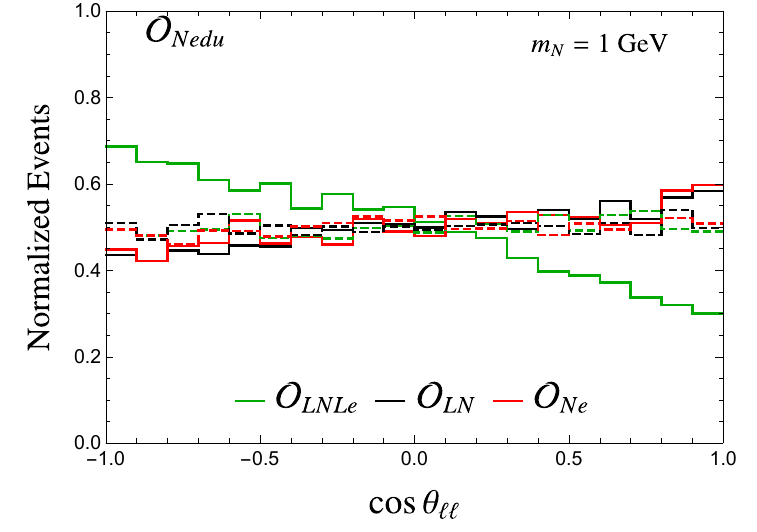}
	\label{fig:cut_smear_ll_cos_nedu_1GeV_eee}
\end{subfigure}
\begin{subfigure}{.32\textwidth}
	\includegraphics[width=\textwidth]{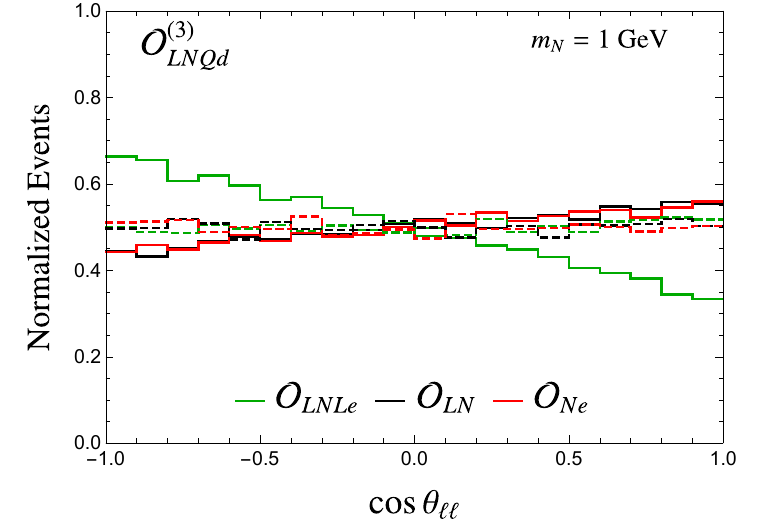}
	\label{fig:cut_smear_ll_cos_lnqd3_1GeV_eee}
\end{subfigure}
\begin{subfigure}{.32\textwidth}
	\includegraphics[width=\textwidth]{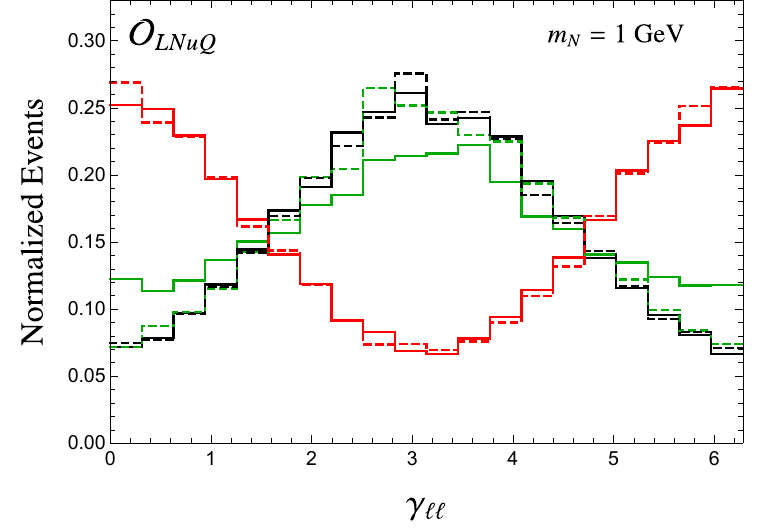}
	\label{fig:cut_smear_ll_ga_lnuq_1GeV_eee}
\end{subfigure}
\begin{subfigure}{.32\textwidth}
	\includegraphics[width=\textwidth]{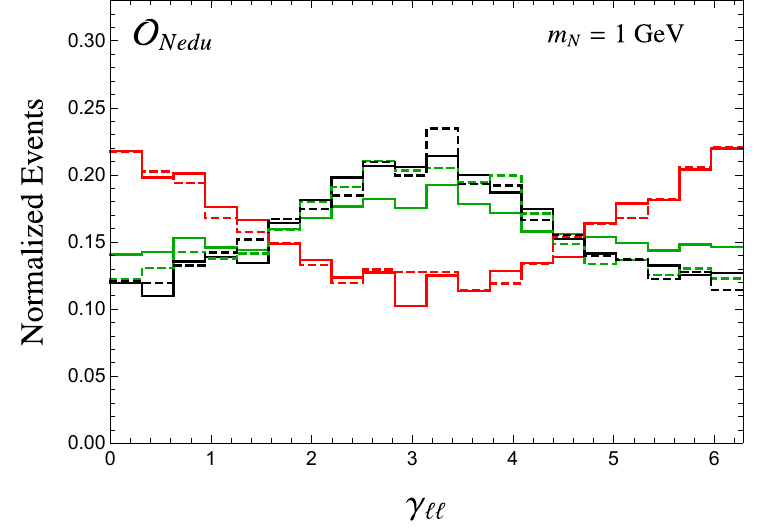}
	\label{fig:cut_smear_ll_ga_nedu_1GeV_eee}
\end{subfigure}
\begin{subfigure}{.32\textwidth}
	\includegraphics[width=\textwidth]{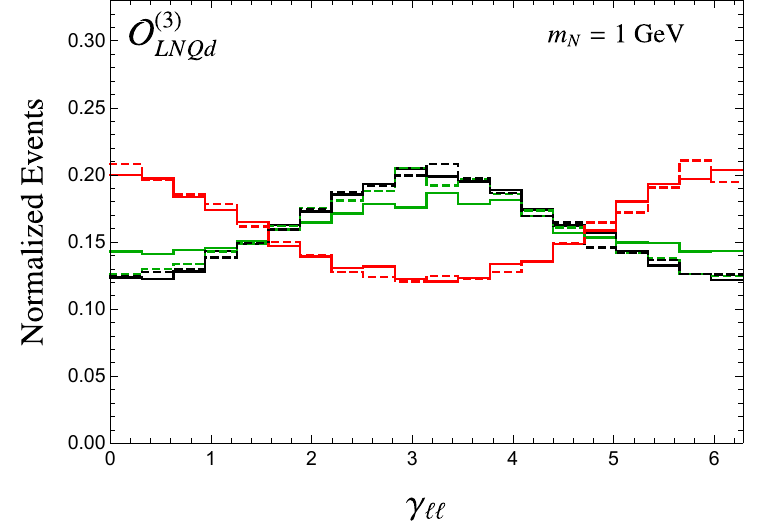}
	\label{fig:cut_smear_ll_ga_lnqd3_1GeV_eee}
\end{subfigure}
\begin{subfigure}{.32\textwidth}
	\includegraphics[width=\textwidth]{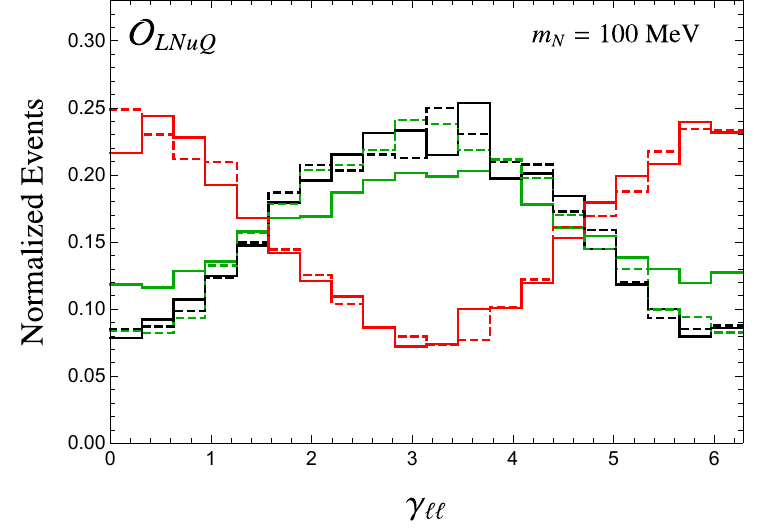}
	\label{fig:cut_smear_ll_ga_lnuq_100MeV_eee}
\end{subfigure}
\begin{subfigure}{.32\textwidth}
	\includegraphics[width=\textwidth]{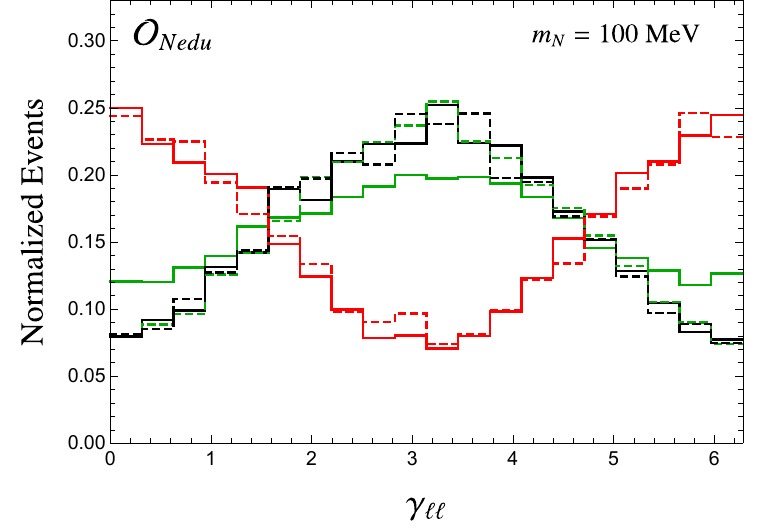}
	\label{fig:cut_smear_ll_ga_nedu_100MeV_eee}
\end{subfigure}
\begin{subfigure}{.32\textwidth}
	\includegraphics[width=\textwidth]{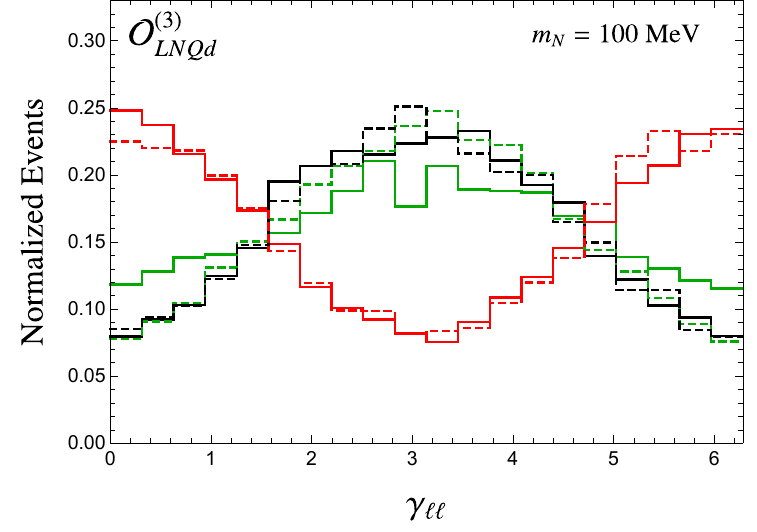}
	\label{fig:cut_smear_ll_ga_lnqd3_100MeV_eee}
\end{subfigure}
	\caption{Smeared $\cos \theta_{\ell\ell}$ distributions~(upper panels)  and $\gamma_{\ell\ell}$ distributions~(middle and lower panels) from three-body $B$ decay for  the three production operators, and three decay operators: $\mathcal{O}_{LNLe}$~(green), $\mathcal{O}_{LN}$~(black), $\mathcal{O}_{Ne}$~(red) for DF~(solid lines) and MF~(dashed lines).
	}
	\label{fig:angle_3body}
\end{figure} 

\begin{table}
	\centering
	\renewcommand{\arraystretch}{1.5}
	\tabcolsep=0.1cm
	\begin{tabular}{|c|c|c|c|c|c|c|c|c|c|c|c|c|}
		\toprule
		Production & \multicolumn{12}{|c|}{$\mathcal{O}_{LNuQ}$} \\
		\hline
		Decay & \multicolumn{4}{|c|}{$\mathcal{O}_{LNLe}$} & \multicolumn{4}{|c|}{$\mathcal{O}_{LN}$}& \multicolumn{4}{|c|}{$\mathcal{O}_{Ne}$} \\
		\hline
		Distribution & $\cos \theta_{\ell\ell}$ & $\gamma_{\ell\ell}$ & $\cos \theta_{\ell\nu}$ &$\gamma_{\ell\nu}$ 
		& $\cos \theta_{\ell\ell}$ & $\gamma_{\ell\ell}$ & $\cos \theta_{\ell\nu}$ &$\gamma_{\ell\nu}$ 
		& $\cos \theta_{\ell\ell}$ & $\gamma_{\ell\ell}$ & $\cos \theta_{\ell\nu}$ &$\gamma_{\ell\nu}$ \\
		\hline
		$m_N = 1$ GeV &174&11&1.3&48  &19&0.3&30&9.0  &26&0.4&39&11          \\
		\midrule
		$m_N = 100$ MeV  &22&9.1&3.0&13  &15&1.5&11&4.8  &21&1.5&22&3.5          \\
		\midrule
		$m_N = 5$ MeV  &4.3&2.9&2.4&2.8  &7.5&1.9&8.2&1.8  &4.6&1.1&2.7&1.6        \\
		
		\toprule
		Production & \multicolumn{12}{|c|}{$\mathcal{O}_{Nedu}$} \\
		\hline
		Decay & \multicolumn{4}{|c|}{$\mathcal{O}_{LNLe}$} & \multicolumn{4}{|c|}{$\mathcal{O}_{LN}$}& \multicolumn{4}{|c|}{$\mathcal{O}_{Ne}$} \\
		\hline
		Distribution & $\cos \theta_{\ell\ell}$ & $\gamma_{\ell\ell}$ & $\cos \theta_{\ell\nu}$ &$\gamma_{\ell\nu}$ 
		& $\cos \theta_{\ell\ell}$ & $\gamma_{\ell\ell}$ & $\cos \theta_{\ell\nu}$ &$\gamma_{\ell\nu}$ 
		& $\cos \theta_{\ell\ell}$ & $\gamma_{\ell\ell}$ & $\cos \theta_{\ell\nu}$ &$\gamma_{\ell\nu}$ \\
		\hline
		  $m_N = 1$ GeV &42&3.1&1.1&16.1  &5.4&0.8&6.4&2.7  &3.7&0.9&6.9&1.9             \\
		\midrule
		$m_N = 100$ MeV  &26&11&4.0&17  &18&1.1&14&3.6  &12&1.3&12&2.8            \\
		\midrule
		$m_N = 5$ MeV  &6.8&1.6&4.6&1.5  &4.3&2.6&3.4&2.6  &5.5&1.5&4.0&0.9          \\
		\toprule
		
		Production & \multicolumn{12}{|c|}{$O^{(3)}_{LNQd}$} \\
		\hline
		Decay & \multicolumn{4}{|c|}{$\mathcal{O}_{LNLe}$} & \multicolumn{4}{|c|}{$\mathcal{O}_{LN}$}& \multicolumn{4}{|c|}{$\mathcal{O}_{Ne}$} \\
		\hline
		Distribution & $\cos \theta_{\ell\ell}$ & $\gamma_{\ell\ell}$ & $\cos \theta_{\ell\nu}$ &$\gamma_{\ell\nu}$ 
		& $\cos \theta_{\ell\ell}$ & $\gamma_{\ell\ell}$ & $\cos \theta_{\ell\nu}$ &$\gamma_{\ell\nu}$ 
		& $\cos \theta_{\ell\ell}$ & $\gamma_{\ell\ell}$ & $\cos \theta_{\ell\nu}$ &$\gamma_{\ell\nu}$ \\
		\hline
		  $m_N = 1$ GeV &30&1.6&0.7&9.1  &3.7&0.2&3.7&1.4  &3.9&0.3&3.8&2.0             \\
		\midrule
		$m_N = 100$ MeV  &18&11&2.7&13  &11&1.1&7.3&2.4  &12&1.1&12&2.3             \\
		\midrule
		$m_N = 5$ MeV  &8.6&2.5&5.1&2.6  &8.1&1.4&5.7&1.3  &5.2&1.1&3.9&1.2           \\
		\toprule
	\end{tabular}
	\caption{Same as Table~\ref{Table: 2body}, except for three-body $B$ decay.}
	\label{Table: 3body}
\end{table}

\begin{figure}[tb]
	\centering
	\begin{subfigure}{.32\textwidth}
		\includegraphics[width=\textwidth]{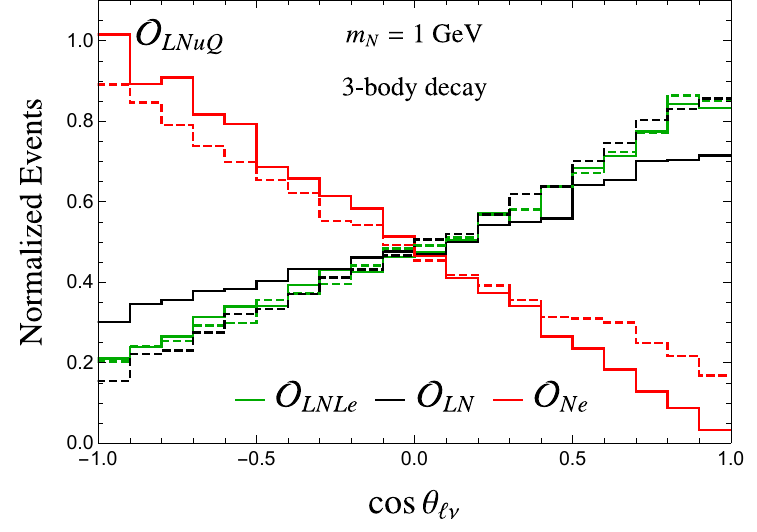}
		\label{fig:cut_smear_ln_cos_lnuq_1GeV_eee}
	\end{subfigure}
	\begin{subfigure}{.32\textwidth}
		\includegraphics[width=\textwidth]{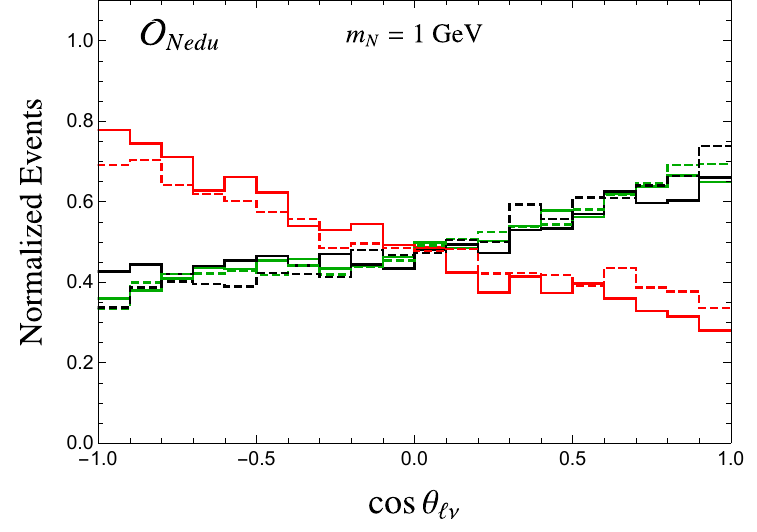}
		\label{fig:cut_smear_ln_cos_nedu_1GeV_eee}
	\end{subfigure}
	\begin{subfigure}{.32\textwidth}
		\includegraphics[width=\textwidth]{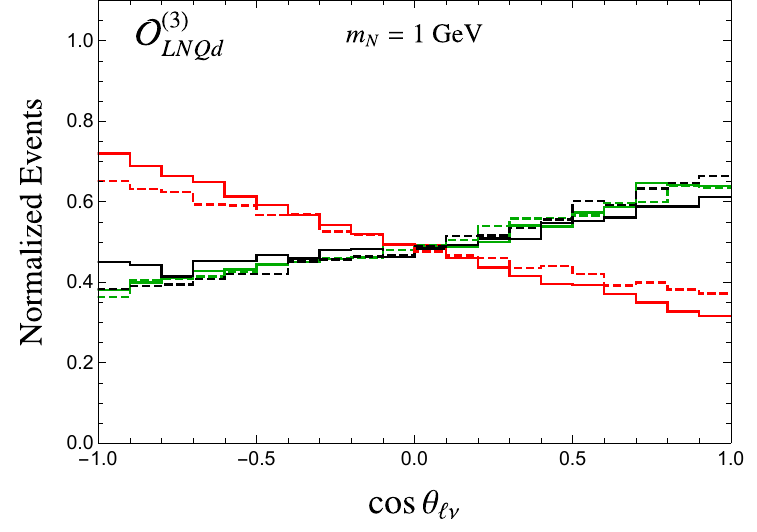}
		\label{fig:cut_smear_ln_cos_lnqd3_1GeV_eee}
	\end{subfigure}
	\begin{subfigure}{.32\textwidth}
		\includegraphics[width=\textwidth]{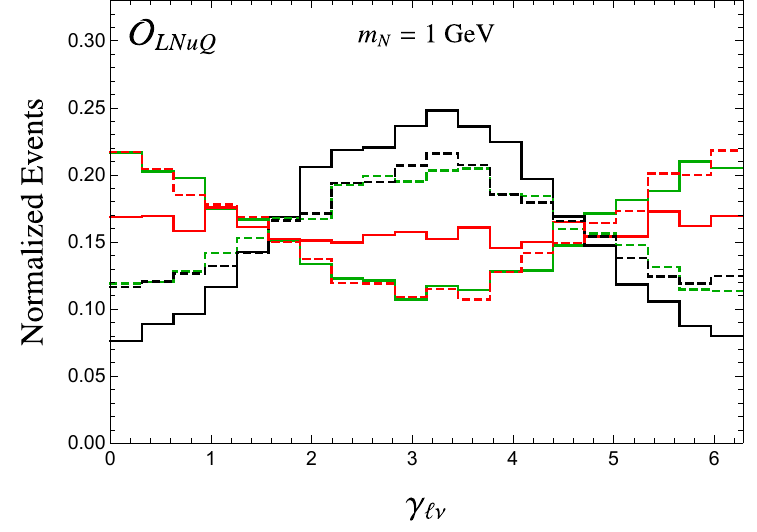}
		\label{fig:cut_smear_ln_ga_lnuq_1GeV_eee}
	\end{subfigure}
	\begin{subfigure}{.32\textwidth}
		\includegraphics[width=\textwidth]{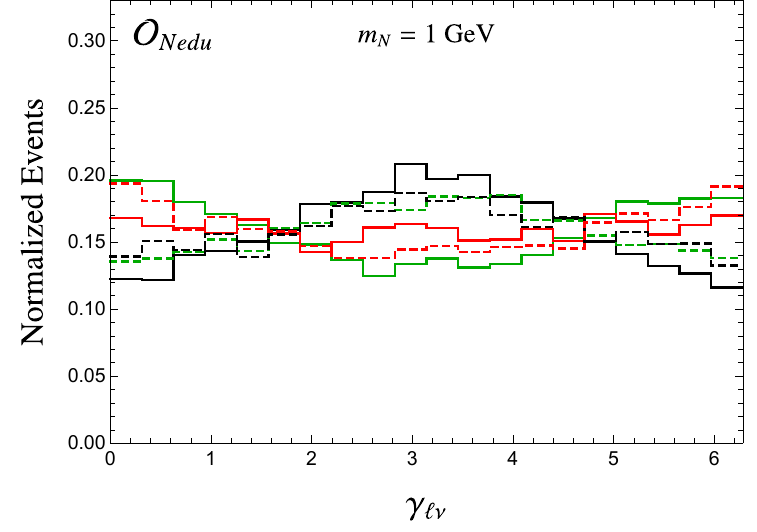}
		\label{fig:cut_smear_ln_ga_nedu_1GeV_eee}
	\end{subfigure}
	\begin{subfigure}{.32\textwidth}
		\includegraphics[width=\textwidth]{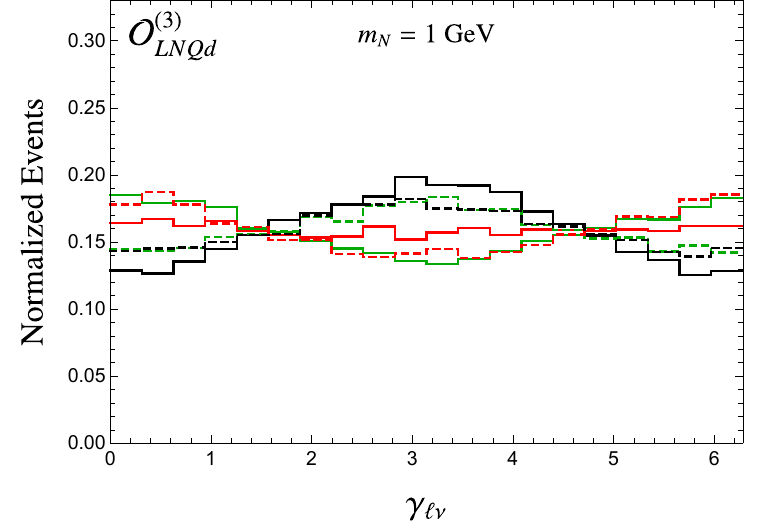}
		\label{fig:cut_smear_ln_ga_lnqd3_1GeV_eee}
	\end{subfigure}
	\begin{subfigure}{.32\textwidth}
		\includegraphics[width=\textwidth]{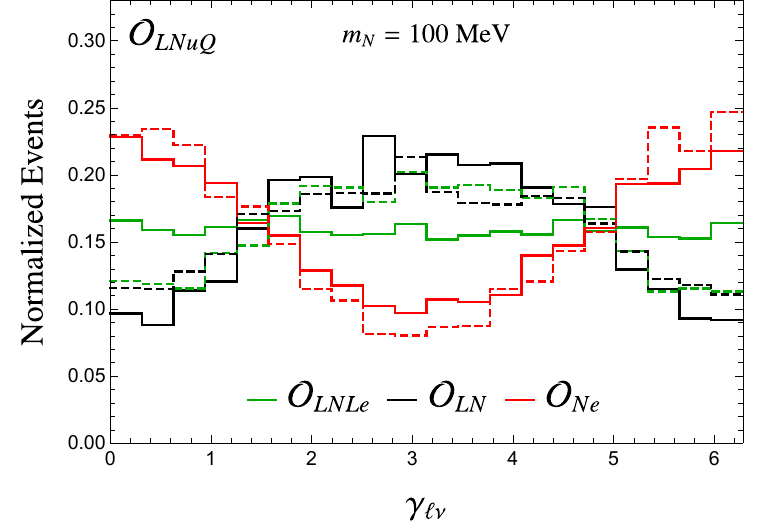}
		\label{fig:cut_smear_ln_ga_lnuq_100MeV_eee}
	\end{subfigure}
	\begin{subfigure}{.32\textwidth}
		\includegraphics[width=\textwidth]{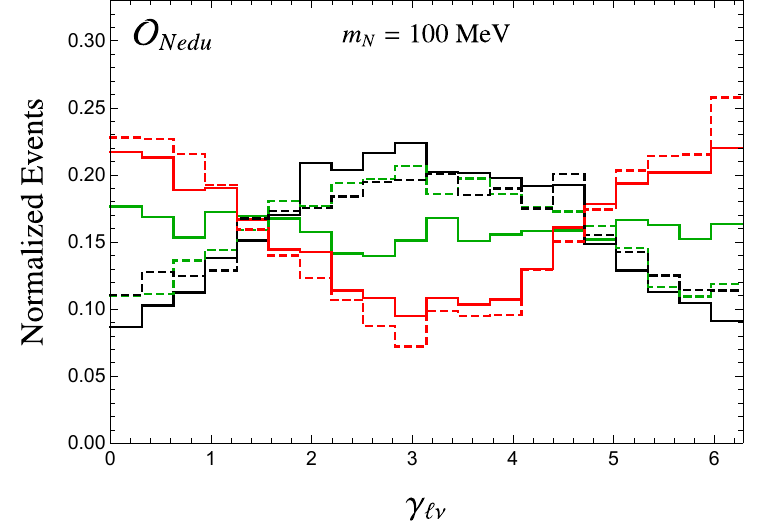}
		\label{fig:cut_smear_ln_ga_nedu_100MeV_eee}
	\end{subfigure}
	\begin{subfigure}{.32\textwidth}
		\includegraphics[width=\textwidth]{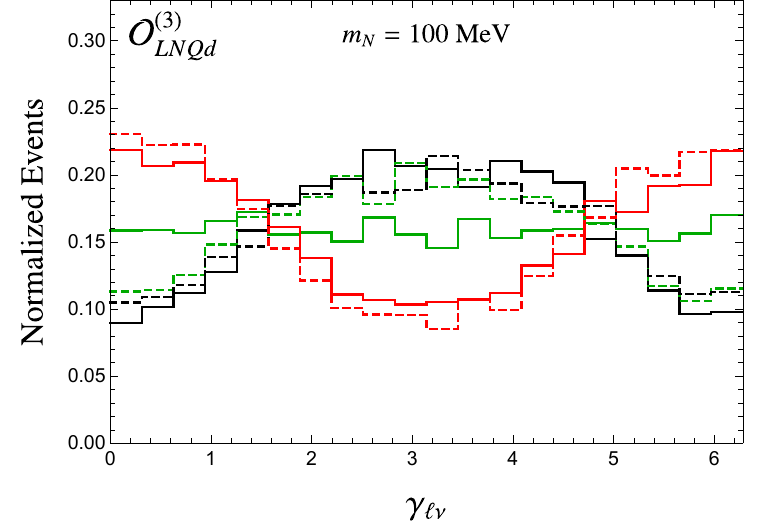}
		\label{fig:cut_smear_ln_ga_lnqd3_100MeV_eee}
	\end{subfigure}
	\caption{Same as Fig.~\ref{fig:angle_3body}, except for angular observables associated with $\ell\nu$.
	}
	\label{fig:angle_3body_ln}
\end{figure}

\subsubsection{Results from three-body $B$ decay}
\label{sec:results3b}
For semi-leptonic three-body $B$ decay, the $\cos \theta_{\ell\ell}$ and $\gamma_{\ell\ell}$ distributions in the $\bar N$ rest frame are shown in Fig.~\ref{fig:angle_3body}. 
We present the angular distributions for the three production and three decay operators.
For $m_N= 1$~GeV, we show the $\cos\theta_{\ell\ell}$ and $\gamma_{\ell\ell}$ distributions and for $m_N=100$~MeV we show only the  $\gamma_{\ell\ell}$ distribution.
As $m_N$ is decreased below 100~MeV, the distributions become similar to the 2-body case in Fig.~\ref{fig:angle_2body}.

The di-lepton angular distributions in the $\bar N$ rest frame depend on both the $\bar N$ polarization and the decay operators, but the angular information of $\bar N$ is lost in the $\bar N$ rest frame. 
For $m_N=1$ GeV, the sensitivity to $N$'s nature is much larger for $\mathcal{O}_{LNuQ}$ than $\mathcal{O}_{Nedu}$ because the $\bar N$ polarization from three-body $B$ decay is only 0.19 for $\mathcal{O}_{Nedu}$. On the other hand, since the $\bar N$ polarizations from two-body $B$ decay are close to unity, it is easier to establish the nature of $N$ from two-body $B$ decay for $m_N=1$~GeV and $\mathcal{O}_{Nedu}$. The values of $\chi^2_{\rm min}$ for three-body $B$ decay are listed in Table~\ref{Table: 3body}. We see that it is easier to distinguish the nature of the RHN for the production operator $\mathcal{O}_{LNuQ}$ than for the other operators if $m_N=1$~GeV. Also, for the decay operator $\mathcal{O}_{LNLe}$, the $\cos \theta_{\ell\ell}$  distribution provides the most sensitivity, while for the decay operators $\mathcal{O}_{LN}$ and $\mathcal{O}_{Ne}$, the $\cos \theta_{\ell\nu}$ distribution offers the most sensitivity; compare the top-left panels of Figs.~\ref{fig:angle_3body} and~\ref{fig:angle_3body_ln}. For $m_N = 100$~MeV, a $5\sigma$ discrimination can only be made for the case with the production operator $\mathcal{O}_{Nedu}$ and decay operator $O_{LNLe}$. Note that for $m_N = 5$~MeV, even a 3$\sigma$ discrimination is not possible for any combination of operators.%


\subsection{Determining the interaction operators }
It is interesting to know whether the production and decay operators can be determined at Belle~II. We first consider the potential to distinguish the production operators from the angular distributions of $B$ decays, and then study how to resolve the production and decay operators from the angular distributions of $\bar N$ decays by separately considering $N$ to be a Dirac or Majorana fermion.

 \subsubsection{Detecting the production operator using three-body $B$ decay}

\begin{figure}[tb]
	\centering
	\begin{subfigure}{.32\textwidth}
		\includegraphics[width=\textwidth]{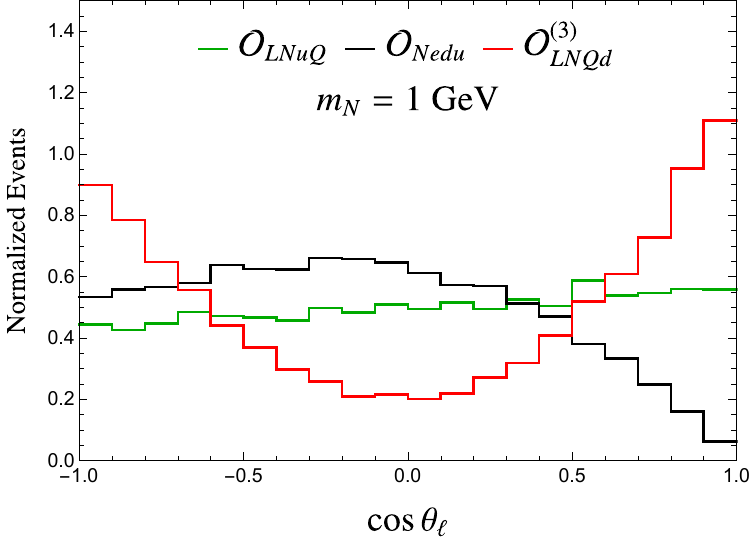}
	\end{subfigure}
	\begin{subfigure}{.32\textwidth}
		\includegraphics[width=\textwidth]{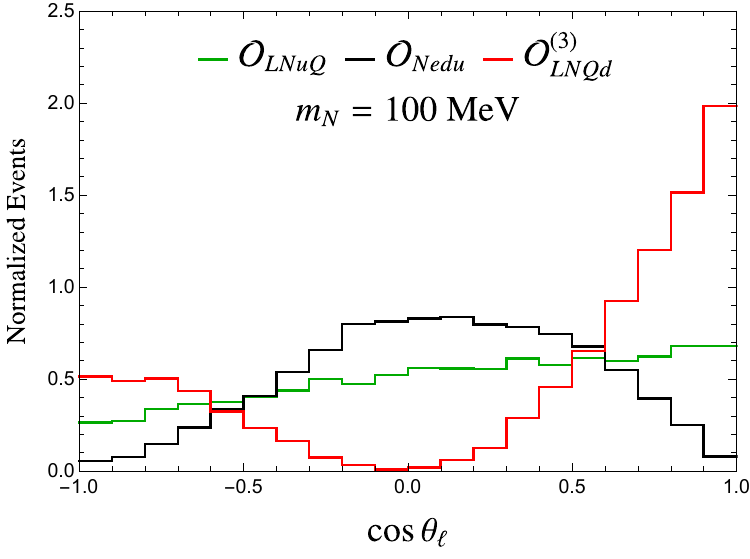}
	\end{subfigure}
	\begin{subfigure}{.32\textwidth}
		\includegraphics[width=\textwidth]{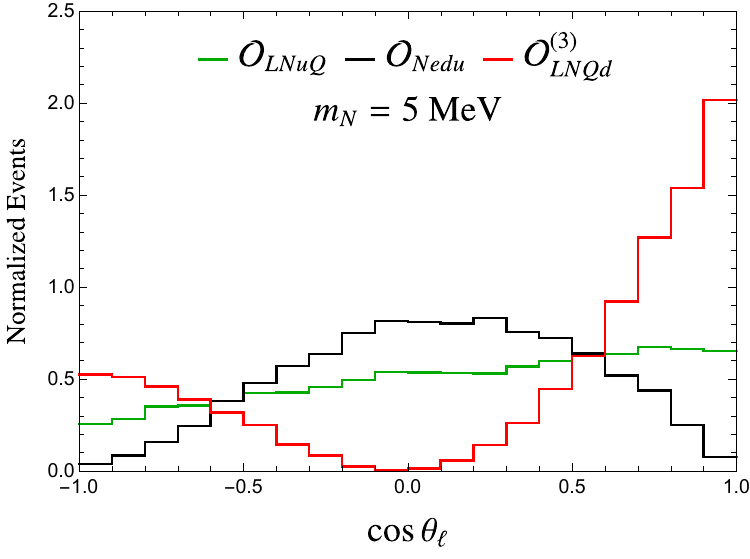}
	\end{subfigure}
	\caption{Same as Fig.~\ref{fig:Nellframe}, but including the smearing and cuts.}
	\label{fig:smear_cosell}
\end{figure}

Figure~\ref{fig:Nellframe} can be used to identify the Lorentz structure of $\bar N$ production operators if $\bar{B}^0 \ra D^{+}e^-\bar N$ and $\bar{B}^0 \ra D^{*+}e^-\bar N$ events can be separated. Here, we only analyze $\bar{B}^0 \ra D^{+}e^-\bar N$ events. To perform an assessment of Belle~II capabilities, we first apply detector smearing and cuts on the final states according to Eqs.~(\ref{eq:smear}) and~(\ref{eq:cut1}). We then boost the charged lepton to the $\ell \bar N$ rest frame and extract the angle between the directions of the charged lepton and $D$ meson. As Fig.~\ref{fig:smear_cosell} shows, the spectral features remain after the smearing and cuts. 

To evaluate the statistical significance of distinguishing between different operators,
we use the $\chi^2$ definition of Eq.~(\ref{eq:chi2}) with $N_{i}^{\rm obs}$ the event counts per $\cos \theta_\ell$ bin simulated with a given production operator. The $\chi^2_{\rm min}$ values for fitting the $\cos \theta_\ell$ distribution with other production operators are listed in Table~\ref{Table: dis_production}. Clearly, all the production operators can be resolved at more than $5\sigma$ regardless of $m_N$. 

\begin{table}
\centering
   \renewcommand{\arraystretch}{1.5}
   \tabcolsep=0.1cm
   \begin{tabular}{|c|c|c|c|c|c|c|}
\toprule
Simulated& \multicolumn{2}{|c|}{$\mathcal{O}_{LNuQ}$} & \multicolumn{2}{|c|}{$\mathcal{O}_{Nedu}$}& \multicolumn{2}{|c|}{$\mathcal{O}^{(3)}_{LNQd}$} \\
\hline
Fitted  & $\mathcal{O}_{Nedu}$ &  $\mathcal{O}^{(3)}_{LNQd}$ 
& $\mathcal{O}_{LNuQ}$ & $\mathcal{O}^{(3)}_{LNQd}$ 
& $\mathcal{O}_{LNuQ}$ & $\mathcal{O}_{Nedu}$  \\
\hline
$m_N = 1$ GeV &124&113&80&288&101&382 \\
\midrule
$m_N = 100$ MeV  &154&602&108&1239&301&864  \\
\midrule
$m_N = 5$ MeV  &164&649&113&1340&296&891      \\
\toprule
\end{tabular}
	\caption{{\it Distinguishing between production operators using three-body $B$ decay.} $\chi^2_{\rm min}$ from the $\cos\theta_\ell$ distributions with different simulated and fitted production operators.  }
	\label{Table: dis_production}
\end{table}

\subsubsection{Detecting the production and decay operators using Dirac $\bar N$ decay}

Figures~\ref{fig:angle_2body} and~\ref{fig:angle_3body} show that the  $\cos\theta_{\ell\ell}$ and $\gamma_{\ell\ell}$  distributions  can be used to ascertain the production and decay operators.  
We first assume that $N$ is a Dirac fermion and that the production and decay  operators are $\mathcal{O}_{LNuQ}$ and $\mathcal{O}_{LNLe}$, respectively, and fit future Belle~II data with the other combinations of production and decay operators. 
We use the $\chi^2$ defined in Eq.~(\ref{eq:chi2}) with $N_{i}^{\rm obs}$ denoting the event counts per bin simulated for Dirac $\bar N$ and $\mathcal{O}_{LNuQ}$ and $\mathcal{O}_{LNLe}$. $\chi^2_{\rm min}$ for different production and decay operators for two-body and three-body $B$ decay are listed in Table~\ref{Table: 2body_interaction} and~\ref{Table: 3body_interaction}, respectively. 

From Table~\ref{Table: 2body_interaction}, it is evident that the sensitivities to resolve different interaction operators becomes smaller as $m_N$ decreases. 
For the two-body $B$ decay mode, the sensitivities to distinguish between the production operators $\mathcal{O}_{LNuQ}$ and $\mathcal{O}_{Nedu}$ are very low. However, for $m_N=1$~GeV, it is possible to distinguish between decay operators at more than 5$\sigma$.
From Table~\ref{Table: 3body_interaction} we see that for the three-body decay mode, the production and decay operators can be distinguished at more than 5$\sigma$ if $m_N=1$~GeV. Also, Belle~II can distinguish between the decay operators at more than 5$\sigma$ for $m_N=100$~MeV, and $\gtrsim 3.5\sigma$  for $m_N=5$~MeV. %
\begin{table}
	\centering
	\renewcommand{\arraystretch}{1.5}
	\tabcolsep=0.1cm
	\begin{tabular}{|c|c|c|c|c|c|c|c|c|c|c|c|c|}
		\toprule
		Production & \multicolumn{12}{|c|}{$\mathcal{O}_{LNuQ}$} \\
		\hline
		Decay & \multicolumn{4}{|c|}{$\mathcal{O}_{LNLe}$} & \multicolumn{4}{|c|}{$\mathcal{O}_{LN}$}& \multicolumn{4}{|c|}{$\mathcal{O}_{Ne}$} \\
		\hline
		Distribution & $\cos \theta_{\ell\ell}$ & $\gamma_{\ell\ell}$ & $\cos \theta_{\ell\nu}$ &$\gamma_{\ell\nu}$ 
		& $\cos \theta_{\ell\ell}$ & $\gamma_{\ell\ell}$ & $\cos \theta_{\ell\nu}$ &$\gamma_{\ell\nu}$ 
		& $\cos \theta_{\ell\ell}$ & $\gamma_{\ell\ell}$ & $\cos \theta_{\ell\nu}$ &$\gamma_{\ell\nu}$ \\
		\hline
		$m_N = 1$ GeV &-&-&-&- &266&15&15&101  &264&119&927&6.5       \\
		\midrule
		$m_N = 100$ MeV  &-&-&-&- &27&6.4&6.8&21  &29&70&107&12         \\
		\midrule
		$m_N = 5$ MeV  &-&-&-&-  &2.5&0.7&2.2&1.2  &4.3&3.3&6.5&2.2       \\
		
		\toprule
		Production & \multicolumn{12}{|c|}{$\mathcal{O}_{Nedu}$} \\
		\hline
		Decay & \multicolumn{4}{|c|}{$\mathcal{O}_{LNLe}$} & \multicolumn{4}{|c|}{$\mathcal{O}_{LN}$}& \multicolumn{4}{|c|}{$\mathcal{O}_{Ne}$} \\
		\hline
		Distribution & $\cos \theta_{\ell\ell}$ & $\gamma_{\ell\ell}$ & $\cos \theta_{\ell\nu}$ &$\gamma_{\ell\nu}$ 
		& $\cos \theta_{\ell\ell}$ & $\gamma_{\ell\ell}$ & $\cos \theta_{\ell\nu}$ &$\gamma_{\ell\nu}$ 
		& $\cos \theta_{\ell\ell}$ & $\gamma_{\ell\ell}$ & $\cos \theta_{\ell\nu}$ &$\gamma_{\ell\nu}$ \\
		\hline
		$m_N = 1$ GeV &0.6&0.3&0.5&0.3  &261&13&17&99  &257&120&927&5.9        \\
		\midrule
		$m_N = 100$ MeV  &1.3&0.4&2.3&0.6  &30&7.3&7.5&21  &34&67&112&11           \\
		\midrule
		$m_N = 5$ MeV  &1.2&0.4&0.8&0.4  &3.9&0.8&2.9&1.4  &4.4&3.6&8.1&1.8         \\
		\toprule
	\end{tabular}
 	\caption{{\it Determining the production and decay operators using $\bar N$ decay.} $\chi^2_{\rm min}$ for two-body $B$ decay  with different production and decay operators. Here, the RHN is a Dirac fermion and is produced (decays) via $\mathcal{O}_{LNuQ}$ ($\mathcal{O}_{LNLe}$). }
 	\label{Table: 2body_interaction}
 \end{table}
\begin{table}
	\centering
	\renewcommand{\arraystretch}{1.5}
	\tabcolsep=0.1cm
	\begin{tabular}{|c|c|c|c|c|c|c|c|c|c|c|c|c|}
		\toprule
		Production & \multicolumn{12}{|c|}{$\mathcal{O}_{LNuQ}$} \\
		\hline
		Decay & \multicolumn{4}{|c|}{$\mathcal{O}_{LNLe}$} & \multicolumn{4}{|c|}{$\mathcal{O}_{LN}$}& \multicolumn{4}{|c|}{$\mathcal{O}_{Ne}$} \\
		\hline
		Distribution & $\cos \theta_{\ell\ell}$ & $\gamma_{\ell\ell}$ & $\cos \theta_{\ell\nu}$ &$\gamma_{\ell\nu}$ 
		& $\cos \theta_{\ell\ell}$ & $\gamma_{\ell\ell}$ & $\cos \theta_{\ell\nu}$ &$\gamma_{\ell\nu}$ 
		& $\cos \theta_{\ell\ell}$ & $\gamma_{\ell\ell}$ & $\cos \theta_{\ell\nu}$ &$\gamma_{\ell\nu}$ \\
		\hline
		$m_N = 1$ GeV &-&-&-&-  &287&13&14&104  &301&127&1005&10          \\
		\midrule
		$m_N = 100$ MeV  &-&-&-&-  &61&9.9&14&28  &67&91&236&21           \\
		\midrule
		$m_N = 5$ MeV  &-&-&-&-  &14&3.1&9.1&3.8  &12&9.3&16&4.9      \\
		
		\toprule
		Production & \multicolumn{12}{|c|}{$\mathcal{O}_{Nedu}$} \\
		\hline
		Decay & \multicolumn{4}{|c|}{$\mathcal{O}_{LNLe}$} & \multicolumn{4}{|c|}{$\mathcal{O}_{LN}$}& \multicolumn{4}{|c|}{$\mathcal{O}_{Ne}$} \\
		\hline
		Distribution & $\cos \theta_{\ell\ell}$ & $\gamma_{\ell\ell}$ & $\cos \theta_{\ell\nu}$ &$\gamma_{\ell\nu}$ 
		& $\cos \theta_{\ell\ell}$ & $\gamma_{\ell\ell}$ & $\cos \theta_{\ell\nu}$ &$\gamma_{\ell\nu}$ 
		& $\cos \theta_{\ell\ell}$ & $\gamma_{\ell\ell}$ & $\cos \theta_{\ell\nu}$ &$\gamma_{\ell\nu}$ \\
		\hline
		  $m_N = 1$ GeV &52&5.6&32&2.3  &220&0.9&46&44  &210&55&304&11           \\
		\midrule
		$m_N = 100$ MeV  &3.3&0.2&3.3&0.8  &65&9.9&17&26  &55&99&208&20            \\
		\midrule
		$m_N = 5$ MeV &3.6&1.5&4.4&1.2  &7.8&1.6&6.3&2.4  &12&8.9&12&5.9         \\
		\toprule
		
		Production & \multicolumn{12}{|c|}{$O^{(3)}_{LNQd}$} \\
		\hline
		Decay & \multicolumn{4}{|c|}{$\mathcal{O}_{LNLe}$} & \multicolumn{4}{|c|}{$\mathcal{O}_{LN}$}& \multicolumn{4}{|c|}{$\mathcal{O}_{Ne}$} \\
		\hline
		Distribution & $\cos \theta_{\ell\ell}$ & $\gamma_{\ell\ell}$ & $\cos \theta_{\ell\nu}$ &$\gamma_{\ell\nu}$ 
		& $\cos \theta_{\ell\ell}$ & $\gamma_{\ell\ell}$ & $\cos \theta_{\ell\nu}$ &$\gamma_{\ell\nu}$ 
		& $\cos \theta_{\ell\ell}$ & $\gamma_{\ell\ell}$ & $\cos \theta_{\ell\nu}$ &$\gamma_{\ell\nu}$ \\
		\hline
		  $m_N = 1$ GeV &64&5.3&36&4.3  &208&1.1&53&38  &205&45&265&12             \\
		\midrule
		$m_N = 100$ MeV  &2.4&1.1&2.8&0.5  &62&9.0&16&25  &58&89&206&19            \\
		\midrule
		$m_N = 5$ MeV  &1.8&1.3&2.2&0.9  &16&2.0&10&2.5  &13&7.7&17&5.0           \\
		\toprule
	\end{tabular}
	\caption{Same as Table~\ref{Table: 2body_interaction}, except for three-body $B$ decay.}
	\label{Table: 3body_interaction}
\end{table}



\subsubsection{Detecting the production and decay operators using Majorana $N$ decay}
If $N$ is a Majorana fermion, the $\cos\theta_{\ell\ell}$ distributions (without smearing) are flat. However, the 
$\cos\theta_{\ell\nu}$ and $\gamma_{\ell\ell}$ distributions are not necessary flat. 
We now assume that $N$ is a MF, and simulate the Belle~II data with the production operator $\mathcal{O}_{LNuQ}$ and decay operator $\mathcal{O}_{LNLE}$, and fit with the other combinations of production and decay operators. The corresponding $\chi^2_{\rm min}$ values  for  two-body and three-body $B$ decay are listed in Table~\ref{Table: 2body_interaction_majorana} and~\ref{Table: 3body_interaction_majorana}, respectively. 
From Table~\ref{Table: 2body_interaction_majorana}, we see that 
it is difficult to distinguish the production and decay operators using the two-body decay mode. From Table~\ref{Table: 3body_interaction_majorana}, we see that in the three-body decay mode, the $\cos\theta_{\ell\nu}$ distribution can be used 
to detect the production operators at more than 5$\sigma$ if $m_N=1$~GeV. .

\begin{table}
	\centering
	\renewcommand{\arraystretch}{1.5}
	\tabcolsep=0.1cm
	\begin{tabular}{|c|c|c|c|c|c|c|c|c|c|c|c|c|}
		\toprule
		Production & \multicolumn{12}{|c|}{$\mathcal{O}_{LNuQ}$} \\
		\hline
		Decay & \multicolumn{4}{|c|}{$\mathcal{O}_{LNLe}$} & \multicolumn{4}{|c|}{$\mathcal{O}_{LN}$}& \multicolumn{4}{|c|}{$\mathcal{O}_{Ne}$} \\
		\hline
		Distribution & $\cos \theta_{\ell\ell}$ & $\gamma_{\ell\ell}$ & $\cos \theta_{\ell\nu}$ &$\gamma_{\ell\nu}$ 
		& $\cos \theta_{\ell\ell}$ & $\gamma_{\ell\ell}$ & $\cos \theta_{\ell\nu}$ &$\gamma_{\ell\nu}$ 
		& $\cos \theta_{\ell\ell}$ & $\gamma_{\ell\ell}$ & $\cos \theta_{\ell\nu}$ &$\gamma_{\ell\nu}$ \\
		\hline
		$m_N = 1$ GeV &-&-&-&-  &0.7&0.2&0.6&0.2  &0.5&191&453&49         \\
		\midrule
		$m_N = 100$ MeV  &-&-&-&-  &0.9&0.4&1.9&0.7  &1.0&104&57&57           \\
		\midrule
		$m_N = 5$ MeV  &-&-&-&-  &0.9&0.4&1.8&0.6  &0.5&5.7&2.0&4.8         \\
		
		\toprule
		Production & \multicolumn{12}{|c|}{$\mathcal{O}_{Nedu}$} \\
		\hline
		Decay & \multicolumn{4}{|c|}{$\mathcal{O}_{LNLe}$} & \multicolumn{4}{|c|}{$\mathcal{O}_{LN}$}& \multicolumn{4}{|c|}{$\mathcal{O}_{Ne}$} \\
		\hline
		Distribution & $\cos \theta_{\ell\ell}$ & $\gamma_{\ell\ell}$ & $\cos \theta_{\ell\nu}$ &$\gamma_{\ell\nu}$ 
		& $\cos \theta_{\ell\ell}$ & $\gamma_{\ell\ell}$ & $\cos \theta_{\ell\nu}$ &$\gamma_{\ell\nu}$ 
		& $\cos \theta_{\ell\ell}$ & $\gamma_{\ell\ell}$ & $\cos \theta_{\ell\nu}$ &$\gamma_{\ell\nu}$ \\
		\hline
		$m_N = 1$ GeV &0.5&0.1&0.4&0.2  &1.1&0.3&0.6&0.3  &0.6&198&461&53         \\
		\midrule
		$m_N = 100$ MeV  &1.4&0.4&1.4&0.5  &1.1&0.3&1.4&0.5  &1.1&105&62&55           \\
		\midrule
		$m_N = 5$ MeV  &1.1&0.6&1.6&0.5  &1.1&0.7&1.2&0.6  &0.6&4.5&2.4&3.5           \\
		\toprule
	\end{tabular}
	\caption{{\it Determining the production and decay operators using $\bar N$ decay.} $\chi^2_{\rm min}$ for two-body $B$ decay  with different production and decay operators. Here, the RHN is a Majorana fermion and is produced (decays) via $\mathcal{O}_{LNuQ}$ ($\mathcal{O}_{LNLe}$). }
	\label{Table: 2body_interaction_majorana}
\end{table}

\begin{table}
	\centering
	\renewcommand{\arraystretch}{1.5}
	\tabcolsep=0.1cm
	\begin{tabular}{|c|c|c|c|c|c|c|c|c|c|c|c|c|}
		\toprule
		Production & \multicolumn{12}{|c|}{$\mathcal{O}_{LNuQ}$} \\
		\hline
		Decay & \multicolumn{4}{|c|}{$\mathcal{O}_{LNLe}$} & \multicolumn{4}{|c|}{$\mathcal{O}_{LN}$}& \multicolumn{4}{|c|}{$\mathcal{O}_{Ne}$} \\
		\hline
		Distribution & $\cos \theta_{\ell\ell}$ & $\gamma_{\ell\ell}$ & $\cos \theta_{\ell\nu}$ &$\gamma_{\ell\nu}$ 
		& $\cos \theta_{\ell\ell}$ & $\gamma_{\ell\ell}$ & $\cos \theta_{\ell\nu}$ &$\gamma_{\ell\nu}$ 
		& $\cos \theta_{\ell\ell}$ & $\gamma_{\ell\ell}$ & $\cos \theta_{\ell\nu}$ &$\gamma_{\ell\nu}$ \\
		\hline
		$m_N = 1$ GeV &-&-&-&-  &0.9&0.8&2.4&0.5  &1.8&203&502&49          \\
		\midrule
		$m_N = 100$ MeV  &-&-&-&-  &3.3&1.0&1.9&0.8  &3.1&153&106&91             \\
		\midrule
		$m_N = 5$ MeV  &-&-&-&-  &2.2&1.0&1.7&1.0  &3.7&20&6.1&14      \\
		
		\toprule
		Production & \multicolumn{12}{|c|}{$\mathcal{O}_{Nedu}$} \\
		\hline
		Decay & \multicolumn{4}{|c|}{$\mathcal{O}_{LNLe}$} & \multicolumn{4}{|c|}{$\mathcal{O}_{LN}$}& \multicolumn{4}{|c|}{$\mathcal{O}_{Ne}$} \\
		\hline
		Distribution & $\cos \theta_{\ell\ell}$ & $\gamma_{\ell\ell}$ & $\cos \theta_{\ell\nu}$ &$\gamma_{\ell\nu}$ 
		& $\cos \theta_{\ell\ell}$ & $\gamma_{\ell\ell}$ & $\cos \theta_{\ell\nu}$ &$\gamma_{\ell\nu}$ 
		& $\cos \theta_{\ell\ell}$ & $\gamma_{\ell\ell}$ & $\cos \theta_{\ell\nu}$ &$\gamma_{\ell\nu}$ \\
		\hline
		 $m_N = 1$ GeV &1.4&14&33&2.9  &0.9&12&28&3.9  &0.6&98&240&23             \\
		\midrule
		$m_N = 100$ MeV  &1.9&0.7&4.0&1.1  &2.4&1.0&3.1&1.1  &3.2&145&117&87              \\
		\midrule
		$m_N = 5$ MeV &3.1&1.4&2.8&0.9  &3.4&1.2&2.1&1.4  &1.2&18&2.1&14          \\
		\toprule
		
		Production & \multicolumn{12}{|c|}{$O^{(3)}_{LNQd}$} \\
		\hline
		Decay & \multicolumn{4}{|c|}{$\mathcal{O}_{LNLe}$} & \multicolumn{4}{|c|}{$\mathcal{O}_{LN}$}& \multicolumn{4}{|c|}{$\mathcal{O}_{Ne}$} \\
		\hline
		Distribution & $\cos \theta_{\ell\ell}$ & $\gamma_{\ell\ell}$ & $\cos \theta_{\ell\nu}$ &$\gamma_{\ell\nu}$ 
		& $\cos \theta_{\ell\ell}$ & $\gamma_{\ell\ell}$ & $\cos \theta_{\ell\nu}$ &$\gamma_{\ell\nu}$ 
		& $\cos \theta_{\ell\ell}$ & $\gamma_{\ell\ell}$ & $\cos \theta_{\ell\nu}$ &$\gamma_{\ell\nu}$ \\
		\hline
		  $m_N = 1$ GeV&0.7&17&40&4.0  &0.9&15&37&3.8  &1.2&94&218&24              \\
		\midrule
		$m_N = 100$ MeV  &2.7&0.6&2.1&0.8  &3.5&1.0&3.0&1.0  &4.2&142&118&77              \\
		\midrule
		$m_N = 5$ MeV  &3.5&1.3&2.0&1.5  &2.8&1.2&1.3&1.3  &1.8&17&4.7&14           \\
		\toprule
	\end{tabular}
	\caption{Same as Table~\ref{Table: 2body_interaction_majorana}, except for three-body $B$ decay.}
	\label{Table: 3body_interaction_majorana}
\end{table}

\section{Summary}
\label{sec:sum}
We performed a detailed study  of the signatures of a RHN $N$ produced in $B$ decay at the Belle~II experiment.
We considered three (scalar, vector and tensor) production operators and three decay operators in the SMNEFT framework. 
	We first showed that, if kinematically accessible, the RHN could be observed by constructing the recoil mass peak at $m_N$.
It is not possible to use lepton number violation to tell if the RHN is a Dirac or Majorana fermion if it decays to same-flavor di-lepton pairs because of the undetected light neutrino in the final state.
However, the angular distributions of the lepton pairs in the RHN rest frame are different for Dirac and  Majorana $N$.
%
We carried out a detailed simulation of the angular distributions of lepton pairs from both two-body and three-body $B$ decay at Belle~II.
The potential to distinguish the nature of the RHN at Belle~II depends on the structure of production and decay operators, and the choice of angular distributions. 
Our results can be summarized as follows.




\begin{enumerate}
	\item  If $N$ is kinematically accessible and decays promptly in the detector, Belle~II has a high discovery potential with a negligible background. Due to the well constrained kinematics, various angular distributions including $\cos\theta_{\ell}$, $\cos\theta_{\ell\ell}$, $\cos\theta_{\ell\nu}$, $\gamma_{\ell\ell}$ and $\gamma_{\ell\nu}$ can be precisely measured at Belle~II, and can be used to determine the MF versus DF nature, and the RHN production and decay operators.

	\item The $\bar{N}$ polarization has a large impact on the di-lepton angular distributions  from $\bar{N}$ decay. We find that the polarization of $\bar N$ is close to 100\% for both
	 $\bar N$ production operators in two-body $B$ decay. The angular distributions of the production operator $O_{LNuQ}$ are very similar to those of $O_{Nedu}$; the tensor production operator $\mathcal{O}^{(3)}_{LNQd}$ does not contribute to two-body $B$ decay. In three-body $B$ decay, it is difficult to distinguish the Dirac versus Majorana nature of a 1~GeV RHN for $O_{Nedu}$ due to the small polarization of $\bar N$ in this case. 
	
	\item $\cos\theta_{\ell\ell}$ and $\cos\theta_{\ell\nu}$ are the best observables to distinguish between DF and MF. $\gamma_{\ell\ell}$ can be important in detecting the $\bar N$ decay operators especially for an MeV RHN. $\gamma_{\ell\ell}$ distributions cannot be used to distinguish between DF and MF for all the $\bar N$ decay operators except $\mathcal{O}_{LNLe}$.	

	\item With 100 signal events, we find that the Dirac versus Majorana nature of the RHN can be distinguished at more than 5$\sigma$ for the two-body and three-body $B$ decay modes for the production operator $O_{LNuQ}$ if $m_N\approx 1$ GeV. 
	The $\cos \theta_{\ell\ell}$ distribution provides the most sensitive probe of the DF/MF question for the decay operator  $\mathcal{O}_{ LNLe}$, while the $\cos \theta_{\ell\nu}$ distribution is the most sensitive probe for the decay operators $\mathcal{O}_{LN}$ and $\mathcal{O}_{Ne}$.    
	
	\item The angular distributions in the $\bar N$ rest frame are strongly affected by detector effects for light $N$. For $m_N\approx 100$~MeV, the Dirac versus Majorana nature can only be established at more than 5$\sigma$ for the case with $O_{Nedu}$ and $O_{LNLe}$ in the three-body $B$ decay mode, and for $m_N \approx 5$~MeV,  only $3\sigma$ sensitivity is achievable for all cases; see Tables~\ref{Table: 2body} and~\ref{Table: 3body}. 
	
	\item The $\cos\theta_{\ell}$ distribution from $\bar B^0 \ra D^+ \ell^- \bar N$ is flat for scalar interactions, and can be used to determine the production operator; see Figs.~\ref{fig:Nellframe} and~\ref{fig:smear_cosell}. With 100 signal events, the production operators can be determined at more than 5$\sigma$ regardless of $m_N$; see Table~\ref{Table: dis_production}.
	 If $\bar N$ is a MF,  the $\cos\theta_{\ell\nu}$ and $\gamma_{\ell\ell}$ distributions are not flat, and can be used to determine the production and decay operators. 
	Note that $\cos\theta_{\ell\ell}$ distributions are flat for MF regardless of the $\bar N$ production operators.

	\item It is difficult to distinguish between production operators with the same decay operator in the two-body decay mode. The decay operator can be determined with huge significance through the $\cos\theta_{\ell\nu}$ distribution for $m_N \gtrsim 100$~MeV. For MF, the $\gamma_{\ell\ell}$ distribution is key to distinguishing between decay and production operators for $m_N \lesssim 100$~MeV; see Tables~\ref{Table: 2body_interaction},~\ref{Table: 3body_interaction},~\ref{Table: 2body_interaction_majorana}, and~\ref{Table: 3body_interaction_majorana}.
\end{enumerate}

\acknowledgments
We thank T.~Browder, B.~Zhang and the rest of the UH Belle II group for many useful inputs.
T.H. was supported in part by the U.S.~Department of Energy under grant 
No.~DE-SC0007914 and in part by the PITT PACC. 
H.L. was supported in part by ISF, BSF and Azrieli foundation.
J.L. was supported by the National Natural Science Foundation of China under Grant No.~11905299 and Guangdong Basic and Applied Basic Research Foundation under Grant No. 2020A1515011479. D.M. was supported in part by the U.S. Department of Energy
under Grant No.~de-sc0010504. D.M. thanks KITP, Santa Barbara for its hospitality, and support via the NSF under Grant No. PHY-1748958, while this work was in progress.

\bibliographystyle{JHEP}
\bibliography{ref}

\end{document}